\documentclass[
aps,
pra,
showpacs
amsmath,
amssymb,
preprint,
floatfix,
lengthcheck, 
superscriptaddress,
longbibliography
]{revtex4-1}

\usepackage{graphicx}
\usepackage{mathtools}
\usepackage{braket}
\usepackage{hyperref}
\usepackage{bbold}
\usepackage{calrsfs}
\usepackage{dsfont}
\usepackage{cancel}
\usepackage{mathrsfs}
\usepackage[normalem]{ulem}

\usepackage{color}

\begin{document}


\title{Multimode equilibrium approximations in light-matter systems\\ from weak to strong coupling
}

\author{Davis M. Welakuh}
\email[Electronic address:\;]{davis.welakuh@mpsd.mpg.de}
\affiliation{Max Planck Institute for the Structure and Dynamics of Matter, Luruper Chaussee 149, 22761 Hamburg, Germany}
\affiliation{Simons Center for Computational Physical Chemistry at New York University, New York, New York 10003, USA}

\author{Vasil Rokaj}
\email[Electronic address:\;]{vasil.rokaj@villanova.edu}
\affiliation{Department of Physics, Villanova University, 
             Villanova, Pennsylvania 19085, USA \looseness=-1}

\author{Michael Ruggenthaler}
\email[Electronic address:\;]{michael.ruggenthaler@mpsd.mpg.de}
\affiliation{Max Planck Institute for the Structure and Dynamics of Matter, Luruper Chaussee 149, 22761 Hamburg, Germany}
\affiliation{The Hamburg Center for Ultrafast Imaging, Luruper Chaussee 149, 22761 Hamburg, Germany}

\author{Angel Rubio}
\email[Electronic address:\;]{angel.rubio@mpsd.mpg.de}
\affiliation{Max Planck Institute for the Structure and Dynamics of Matter, Luruper Chaussee 149, 22761 Hamburg, Germany}
\affiliation{Center for Initiative for Computational Catalysis and Computational Quantum Physics, Flatiron Institute, 162 5th Avenue, New York, NY 10010, USA}


\begin{abstract}
In this work, we detail different approaches to treat multi-mode photonic environments within non-relativistic quantum electrodynamics in the long-wavelength approximation efficiently. Specifically we show that for equilibrium properties of coupled light-matter systems, we can approximately capture the effects of multi-mode photonic environments on matter systems by either only keeping the polarization part of the electric field in the length-gauge formulation or by a few effective modes. We present a comprehensive set of approximation methods designed to accurately capture equilibrium phenomena in quantum light-matter systems across a range of complex photonic environments, from weak to strong coupling. These methods are applied to atomic and molecular models as well as to a two-dimensional quantum ring, demonstrating the versatility of our approach and laying the groundwork for first-principles simulations of real materials in cavity quantum electrodynamics.
\end{abstract}

\maketitle


\section{Introduction}
\label{sec:introduction}

The nascent fields of polaritonic chemistry~\cite{ebbesen2016,ruggenthaler2017b,flick2018a,garcia-vidal2021,ruggenthaler2023,mandal2023,sidler2022} and cavity quantum materials engineering~\cite{schlawin2022,ebbesen2016,garcia-vidal2021, BaydinCavityReview} have achieved tremendous experimental breakthroughs in the modification or control of the chemical and physical properties of complex many-body systems strongly coupled to photons. Some examples of pioneering experiments include the demonstration of polariton lasing~\cite{cohen2010}, control of photochemical reactions~\cite{hutchison2012} and energy transfer~\cite{coles2014,zhong2016}, modification of ground-state chemical reactions via vibrational strong coupling~\cite{thompson2006,thomas2019}, enhancement of harmonic generation from polaritonic states~\cite{chervy2016,barachati2018}, experimental designs for chiral light-matter interactions~\cite{Tay2025, kulkarni2025chiralcavity, hubener2021engineering, AndbergerEtAl2024PRB, AupiaisEtAl2024AP, Suarez-ForeroEtAl2024SA}, and cavity control of condensed matter properties~\cite{peter2005,liu2015, FaistCavityHall, paravacini2019, EnknerPRX2024, keren2025}. Alongside these significant experimental advances, first-principles approaches from quantum chemistry and electronic-structure theory have been extended to ab initio quantum electrodynamics (QED)~\cite{ruggenthaler2014,flick2017b,flick2019,flick2018,welakuh2022,mandal2023,haugland2020,jestaedt2020,buchholz2019,riso2022}. However, the humongous number of degrees of freedom of molecular or solid-state systems strongly interacting with a continuum of photon modes makes the use of approximation strategies necessary for the simulation of real materials from first-principles.

A very common strategy to simplify light-matter coupled problems is to assume that light and matter interact weakly, and use a perturbative treatment~\cite{yassir2025}. Indeed, for certain observables, such as the spontaneous emission rate in free space, perturbation theory is very accurate~\cite{craig1998}. However, in the novel regimes of strong and ultrastrong coupling between light and matter~\cite{kockum2019ultrastrong, MavronaUSCFP, Garziano, BaydinCavityReview}, non-perturbative strategies and approximations become necessary. If photonic observables are the target, a common approximation is to reduce the matter system to a few levels and to keep most of the photon degrees of freedom~\cite{dicke1954,tavis1968,garraway2011}. On the other side, if we aim to understand or determine changes in chemical and material properties, it is necessary to keep all the relevant matter degrees of freedom. 
In this case, the photonic description is simplified and the continuum of modes of a real photonic environment is reduced to a few effective modes~\cite{tannoudji1989,svendsen2023theory}. This approach is common to model certain effects, such as the Rabi splitting that is apparent in an absorption measurement for atomic or molecular systems~\cite{mandal2023}, but it is known to not capture (without further adaptations) specific effects, such as the Purcell enhancement of spontaneous emission for excited states~\cite{purcell1995}. Nevertheless, by reducing the degrees of freedom of the photon field to a few effective ones, we can in practice use a more detailed description of the matter subsystem. It is therefore important to understand under which conditions such a simplification is viable, specifically in the context of ab initio quantum electrodynamics (QED). Recent considerations for extended systems~\cite{svendsen2023theory} and for finite model systems~\cite{welakuh2025} suggest that restricting to a few effective modes is a reasonable approximation strategy to determine photon-induced changes in the equilibrium properties of matter.

In this work, we present a comprehensive toolset of approximation methods designed to accurately capture equilibrium phenomena in quantum light-matter systems across a range of complex multimode photonic environments, from weak to strong coupling. The approximation methods aim to model coupled light-matter systems at equilibrium within non-relativistic QED in the long-wavelength approximation.

We apply these methods to one-dimensional atomic and molecular models and to a two-dimensional quantum ring system coupled to a discretized continuum of photon modes. Specifically, we test three different strategies to define the effective treatment of the photon field: (i) We include only the self-polarization of matter due to the photon field, also known as dipole self-energy. This approximation works well for matter observables under coupling to a relatively small number of photon modes, $N_p\approx10-50$. Above that, as a result of the lack of photon-matter correlations, it  overestimates the effect of photons on matter and deviates strongly from the exact result. (ii) We keep only the highest occupied photon mode, which is the one with the lowest frequency. This approximation performs surprisingly well, as we find that the contributions of the higher-lying modes are rather small, and suppressed due to the energetic difference from the ground-state. (iii) We perform an averaging procedure over all occupied modes and define an effective single-mode field which is then treated fully quantum mechanically. This approximation outperforms the other methods in all test cases. This is a remarkable result, as it suggests that the averaging method can be used for first-principles simulations of realistic systems coupled to complex multi-mode electromagnetic environments.

\section{Theoretical framework}
\label{sec:theory}

Our starting point to describe the interaction of electrons, photons, and nuclei is the non-relativistic limit of quantum electrodynamics (QED)~\cite{spohn2004}. In this setting, we invoke the long-wavelength limit (or dipole approximation)~\cite{tannoudji1989} and already assume a discretized electromagnetic continuum of modes~\cite{rokaj2017,spohn2004,svendsen2023theory}. The length form of the Pauli-Fierz Hamiltonian is used to describe the interaction and is given by
\begin{align} 
\hat{H}_{\text{L}} &=  \sum\limits_{l=1}^{N_{e}} \frac{\hat{\textbf{p}}_{l}^{2}}{2m} + \sum\limits_{l=1}^{N_{n}}\frac{\hat{\textbf{P}}_{l}^{2}}{2M_{l}}   - \sum\limits_{l=1}^{N_{e}} \sum\limits_{j=1}^{N_{n}} Z_{j}w(|\hat{\textbf{r}}_{l} - \hat{\textbf{R}}_{j}|) \nonumber \\
& \quad + \frac{1}{2} \sum\limits_{l\neq j}^{N_{n}}Z_{l}Z_{j}w(|\hat{\textbf{R}}_{l} - \hat{\textbf{R}}_{j}|) + \frac{1}{2} \sum\limits_{l\neq j}^{N_{e}}w(|\hat{\textbf{r}}_{l} - \hat{\textbf{r}}_{j}|)  \nonumber \\
& \quad + \frac{1}{2} \sum_{\alpha=1}^{N_{p}} \left[\hat{p}^2_{\alpha} + \omega^2_{\alpha}\left(\hat{q}_{\alpha} \!-\! \frac{\boldsymbol{\lambda}_{\alpha}}{\omega_{\alpha}} \cdot \hat{\boldsymbol{\mu}}  \right)^2\right] . \label{eq:length-gauge-hamiltonian}
\end{align}
Here, the positive parameters $m$ and $M_l$ are the \textit{bare masses} of the $N_{e}$ electrons and $N_{n}$ nuclei, respectively~\cite{spohn2004,welakuh2025}. The coordinates $\hat{\textbf{r}}_{l}$ and $\hat{\textbf{R}}_{l}$ and the momenta $\hat{\textbf{p}}_{l}$ and $\hat{\textbf{P}}_{l}$, respectively, describe the electrons and nuclei, and $w$ is the longitudinal interaction between the charged particles with charge number $Z_{j}$. The total dipole operator is $\hat{\boldsymbol{\mu}} =-\sum_{l=1}^{N_{e}} |e| \, \hat{\textbf{r}}_{l} +  \sum_{l=1}^{N_{n}}Z_{l}|e|\hat{\textbf{R}}_{l}$ where $e$ is the \emph{observable} electronic charge. In free space and in three dimensions, $w$ is the usual Coulomb interaction $w(|\hat{\textbf{r}} - \hat{\textbf{r}}'|)=e^{2}/4\pi\varepsilon_{0}|\hat{\textbf{r}} - \hat{\textbf{r}}'|$. The energy of the quantized electromagnetic field with mode frequency $\omega_{\alpha}$ for each mode $\alpha$ of an arbitrarily large but finite number of photon modes $N_{p}$ is given in terms of the displacement coordinate $\hat{q}_{\alpha}$ and conjugate momentum $\hat{p}_{\alpha} = -i\hbar \tfrac{\partial}{\partial \hat{q}_{\alpha}}$ operators that satisfy the commutation relation $\left[\hat{q}_{\alpha}, \hat{p}_{\alpha'}\right]= i\hbar\delta_{\alpha,\alpha'}$. The light-matter coupling is represented by the parameter $\boldsymbol{\lambda}_{\alpha}=\sqrt{1/\epsilon_{0}V_{\alpha}}\textbf{e}_{\alpha}$, where $V_{\alpha}$ is the quantization mode volume of the mode $\alpha$. 


The properties of a matter system that explicitly interacts with the modes of an electromagnetic field can be obtained by solving the static Schr\"{o}dinger equation for Eq.~\eqref{eq:length-gauge-hamiltonian}. To capture the details of the coupled system, we do not only need to treat the matter system in detail but also describe the multi-mode electromagnetic field in the same manner. This commonly implies including many (in principle, a continuum of) modes of the electromagnetic field. For realistic light-matter systems, a non-perturbative ab-initio simulation becomes computationally very demanding. To be able to tackle such numerically challenging tasks, first-principles theoretical approaches capable of treating the coupled light-matter system have been developed~\cite{ruggenthaler2014,flick2015,flick2019,welakuh2022,haugland2020,flick2018}. These first-principles approaches correctly account for the multi-mode nature of the electromagnetic environment, but some of them still run into high computational cost both in the weak and strong coupling regimes due to explicitly including the photonic continuum~\cite{haugland2020,flick2019}. In the following section, we present different approximation strategies to capture effects that arise due to the interaction of matter systems with an electromagnetic continuum.

\section{Approximation strategies for multi-mode ground-state modifications}
\label{sec:approximations-to-NRQED}

The approximation strategies presented here focus on capturing the effects of a continuum of modes on the equilibrium properties of the matter subsystem. We can do so because it can be shown that the Pauli-Fierz Hamiltonian has a coupled light-matter ground state for systems that have an uncoupled (bare matter) ground-state~\cite{spohn2004}. Bare excited states, in contrast, turn into resonances and hence obtain a finite life-time. This also highlights that the proposed approximation strategies will not necessarily capture all details of excited-state and time-dependent problems, which can depend crucially on the properties of the continuum of electromagnetic modes~\cite{flick2019, ruggenthaler2023}.

\subsection{Matter plus dipole self-energy}
\label{subsec:QM+DSE}

The first approximation strategy is specific to the length-gauge form of the Pauli-Fierz Hamiltonian. This approximation strategy is simple, as it accounts only for the mean-field contribution of the displacement part of the photon field and merely keeps the polarization terms fully quantum. In this approach, we drop all the terms $\hat{p}_{\alpha}^2$ in Eq.~\eqref{eq:length-gauge-hamiltonian} and simply keep $q_{\alpha} = \langle \hat{q}_{\alpha} \rangle$. Due to the zero-field condition for any eigenstate~\cite{flick2018,schaefer2020} we have\begin{align}
q_{\alpha} = \frac{\boldsymbol{\lambda}_{\alpha}\cdot \langle \hat{\boldsymbol{\mu}} \rangle}{\omega_{\alpha}} . \nonumber
\end{align}
We call the resulting approach matter plus dipole self-energy approximation denoted (M$+$DSE), such that the Hamiltonian becomes
\begin{align}
\hat{H}_{\text{M+DSE}} &= \hat{H}_{\text{M}} 
+ \frac{1}{2} \sum_{\alpha=1}^{N_{p}}\left(\boldsymbol{\lambda}_{\alpha}\cdot \langle \hat{\boldsymbol{\mu}} \rangle \!-\! \boldsymbol{\lambda}_{\alpha} \cdot \hat{\boldsymbol{\mu}}  \right)^2 ,
\label{eq:el-pt-hamiltonian-length-DSE}
\end{align}
where
\begin{align} 
\hat{H}_{\text{M}} &= \sum\limits_{l=1}^{N_{e}}\frac{\hat{\textbf{p}}_{l}^{2}}{2m} + \sum\limits_{l=1}^{N_{n}}\frac{\hat{\textbf{P}}_{l}^{2}}{2M_{l}} - \sum\limits_{l=1}^{N_{e}} \sum\limits_{j=1}^{N_{n}} Z_{j}w(|\hat{\textbf{r}}_{l} - \hat{\textbf{R}}_{j}|) \nonumber \\
& \quad + \frac{1}{2} \sum\limits_{l\neq j}^{N_{e}} w(|\hat{\textbf{r}}_{l} - \hat{\textbf{r}}_{j}|) + \frac{1}{2} \sum\limits_{l\neq j}^{N_{n}}Z_{l}Z_{j}w(|\hat{\textbf{R}}_{l} - \hat{\textbf{R}}_{j}|) \, , \label{eq:matter-hamiltonian}
\end{align}
is the matter Hamiltonian. We note that in Eq.~\eqref{eq:matter-hamiltonian} we use the bare masses of electrons $m$ and nuclei $M_{l}$ given in Eq.~(\ref{eq:length-gauge-hamiltonian}). That is, the photon contribution described via the dipole self-energy term accounts approximately for the photon-mass contribution to the observable mass of the charged particles~\cite{welakuh2025}. In the case of a single or few modes, this approximation strategy is sometimes called the QED-Hartree-Fock~\cite{haugland2020} or cavity-Born-Oppenheimer approximation for the electronic subsystem with zero-field condition~\cite{flick2017b}. 
Since the field is reduced to an effective dipole-dipole (polarization) interaction, the computational complexity reduces to solving the length-gauge (dressed) matter subsystem only. 
We note, however, that the dipole self-energy term is not a bare matter operator. In contrast, as can be seen from Eqs.~\eqref{eq:annahil-length} and \eqref{eq:create-length}, this term arises from the photon-field energy in the original Coulomb (velocity) gauge Hamiltonian of Eq.~\eqref{eq:velocity-gauge-hamiltonian}~\cite{rokaj2017}.

\subsection{Effective few-mode descriptions}
\label{subsec:effective-one-mode}

\begin{figure}[bth]
\includegraphics[width=1.0\columnwidth]{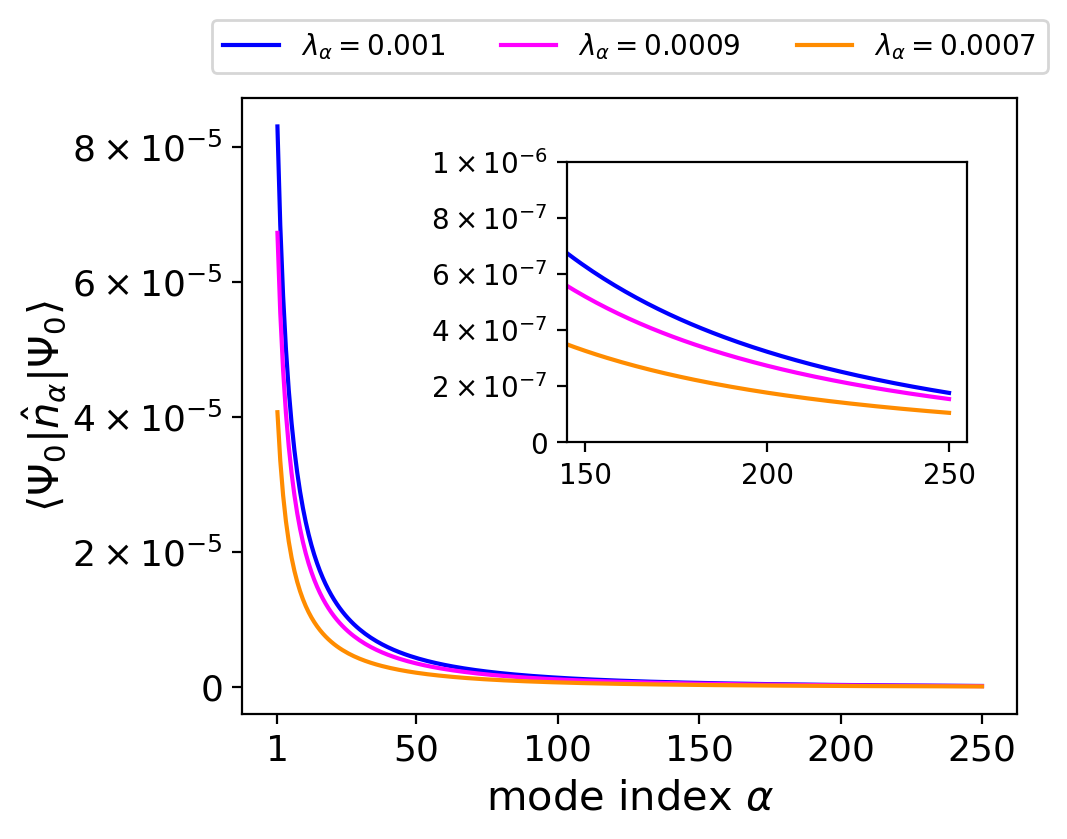}
\caption{The photon occupation of the combined ground-state for the atomic system coupled to a discretized continuum sampled with $N_{p}=250$ photon modes. The lowest modes couple strongest and the occupation increases with the coupling strength. The inset shows the same exponential decrease for higher-lying modes.}
\label{fig:pt-occup-per-mode}
\end{figure}

A different approximation strategy is to replace the continuum of modes by a few effective modes. There are different ways in which one sets up such a few-mode approximation, and which one is most appropriate depends on the coupled system under considerations and also on which observables are targeted. A widely used few-mode approach in the context of polaritonic chemistry and cavity materials engineering~\cite{ruggenthaler2018quantum,hubener2021engineering, DagRokaj2024,ebbesen2016,garcia-vidal2021} is to subsume most of the continuum of modes into the renormalized (observable) masses of the charged particles and to model the enhanced part of the photonic density of states as a few effective modes~\cite{flick2019,svendsen2023theory}. Here we do not follow this strategy, but instead work with the original bare masses and define effective modes that also capture these mass-renormalization effects~\cite{spohn2004,welakuh2025}. Moreover, the focus is on the equilibrium properties of the matter subsystem and 
we do not focus on capturing excited-state properties such as resonances~\cite{flick2019,ruggenthaler2023,welakuh2022a,welakuh2022}. In the following, we give two possible ways to determine these effective modes for the cases that we consider in this work.
\\
\\
\noindent \textbf{Relevant modes:} The first way to define an effective-mode approximation is to keep only those photon modes that are occupied the most.  That is, we only treat those photon modes explicit that couple the strongest to the matter subsystem. In the one-dimensional cases that we consider (see Sec.~\ref{subsubsec:atomic-system} and \ref{subsubsec:molecular-system}), it is relatively easy to determine a priori which are the most relevant photon modes. Since we consider a discretized continuum of modes, where all frequencies have the same fundamental coupling strength $\lambda$, the effective coupling is inversely proportional to the mode frequency. Thus, as can be seen from Fig.~\ref{fig:pt-occup-per-mode}, the lowest-lying photon modes are the most important ones in the one-dimensional cases that we consider. However, this immediately raises the question, specifically for a free-space continuum of modes, where no intrinsic infrared length-scale/cutoff is available, whether we run into an infrared divergence. In the case of non-relativistic quantum electrodynamics (NRQED), the exact analytic theory does not have an infrared-divergence~\cite{spohn2004}. However, handling the frequency-zero limit of the photon field needs extra care. This is also the case for numerical simulations. As discussed in Ref.~\cite{welakuh2025}, even for the long-wavelength approximation, it is important that the matter and photonic discretizations/scales agree. That is, the finite simulation box for the matter subsystem provides a natural infrared cutoff for the photon modes. This effective-mode approximation that retains only the lowest-frequency modes which couple strongly to the bound matter system is denoted as ``NRQED$_\text{low}$''.

We note that the importance of the higher-lying modes depends on the dimensionality of the problem. In two and three dimensions, while they effectively have a smaller interaction strength due to $\lambda/\sqrt{\omega_{\alpha}}$, for higher frequencies, there are many more modes within a small energy window. Indeed, the necessity of a corresponding ultraviolet cutoff, which in our simulations is naturally provided by the grid-point distances (see also App.~\ref{sup:quantum-ring}), is specifically apparent in those higher-dimensional cases~\cite{welakuh2025}. Thus, for two and three dimensions, the effective coupling strength needs to be weighted by the density of photon modes within a small energy range in order to single out the relevant effective modes.
\\
\\
\noindent \textbf{Averaged modes:} A different approach is to use an appropriate averaging over all modes and subsume the effective coupling strength into a few modes. In this way, we also take into account the difference in mode densities in two and three dimensions directly. The number of averaged modes $\tilde{N}_p$ that we consider is at least as large as the real-space dimension of the problem. That is, we assume at least one averaged effective mode for each polarization direction to which the matter system can couple. For the cases of frequency-independent fundamental coupling strengths, we first divide the modes according to their respective polarization directions, that is $N_{x}$ modes we use in the following.
\begin{align} 
\omega_{\tilde{\alpha}} = \frac{\sum_{\alpha=1}^{N_{p}}\omega_{\alpha}}{N_{p}} \, , \quad \textrm{and} \quad \lambda_{\text{ave}}=\sqrt{\sum_{\alpha}\lambda_{\alpha}^2} \, . \label{eq:average-freq-lambda}
\end{align}
This approach is denoted as ``NRQED$_\text{ave}$'' since it averages only over the photon mode frequencies that are valid under the dipole approximation in the setting of non-relativistic QED. The averaging over the photon-mode frequencies can be done as follows. For the case where we sampled 250 modes, the average cavity frequency is $\omega_{\text{ave}}\!=\!0.255$~a.u. (see Sec.~\ref{sec:result-and-discussion}). For an effective single-mode approximation, the computational cost of the combined matter-photon problem is reduced to treating only one photon mode including the matter degrees.

\section{Results and Discussion}
\label{sec:result-and-discussion}

To investigate how well the different approximation strategies capture equilibrium properties of coupled light-matter systems, we consider simple models of bound-state systems that interact with an electromagnetic continuum. Specifically, we consider one-dimensional models of the hydrogen atom and molecule (i.e. H$_{2}$) coupled to a discretized one-dimensional sampling of the electromagnetic continuum. In addition, we consider a two-dimensional model of a semiconductor quantum ring of GaAs coupled to a discretized two-dimensional sampling of the electromagnetic continuum. An approximation strategy is considered to perform well if the results for the observables under investigation qualitatively agree with those obtained from non-perturbative NRQED. The latter are determined through exact diagonalization of the Hamiltonian describing the model system interacting with the discretized electromagnetic continuum. For the atomic system, we compare the approximation strategies to the exact NRQED result by investigating key observables such as the ground-state energy and the ground-state density.
The ground-state energy reflects the overall stability of the system and defines its absolute energetic scale, and the ground-state density encodes the spatial distribution and localization of the electrons. For the molecular system, we compare the approximation strategies to the exact NRQED result by computing the classical dissociation energy. This quantity characterizes how strongly the atoms are bound together and provides direct information about bond strength, stability, and the energy required to separate the fragments. In the case of the atomic and molecular systems, the discretized electromagnetic continuum is constructed as follows. The polarizations of the photon modes of the discretized continuum are along the dimension in which the charged particles are allowed to move. We sample the photon continuum so that the range of its frequencies covers the desired energy range of the bound-state systems. The lower and upper energy cutoffs given in atomic units (a.u.) are, respectively, 0.01 a.u. and 0.5 a.u., while we sample the one-dimensional continuum by explicitly including 250 photon modes with equidistant energy spacing per mode of 0.00197 a.u. . To investigate the approximation strategy where we consider the relevant photon modes (i.e. NRQED$_\text{low}$), we choose the lowest frequency mode $N_{p}\!=\!1$ with cavity frequency $\omega_{\alpha=1}\!=\!0.01$~a.u. out of the 250 modes and the original light-matter coupling. For the coupled light-matter systems, we investigate the influence of the photon continuum on the properties of the matter subsystem.

\subsection{Atom-photon system}
\label{subsubsec:atomic-system}

We start by investigating how the discretized continuum of photon modes interacts with a bound-state system. Specifically, we consider how each of the sampled-photon modes couples to the matter system and influences its properties. We do this for the ground-state of a model of atomic hydrogen (see App.~\ref{sup:1D-atomic-hydrogen} for details) by computing the mean photon occupation per photon mode $n_{\alpha} = \langle \Psi_{0}|\hat{n}_{\alpha}|\Psi_{0}\rangle$ where $|\Psi_{0}\rangle$ is the correlated ground-state of the atom-photon system and $\hat{n}_{\alpha}$ is the photon number operator defined in the length gauge as given in Eq.~(\ref{eq:photon-occupation}). At this point, it is important to note that the photon mode occupation operator is given in terms of the annihilation and creation operators $\hat{n}_{\alpha}=\hat{a}^{\dagger}_{\alpha}\hat{a}_{\alpha}$ where the photon operators in the length form are
\begin{align}
\hat{a}_{\alpha} 
&= \frac{1}{\sqrt{2\hbar \omega_{\alpha}}} \left[\hat{p}_{\alpha} - i\omega_{\alpha} \left(\hat{q}_{\alpha} - \frac{\boldsymbol{\lambda}_{\alpha}\cdot \hat{\boldsymbol{\mu}}}{\omega_{\alpha}}\right) \right] , \label{eq:annahil-length} \\
\hat{a}_{\alpha}^{\dagger} 
&= \frac{1}{\sqrt{2\hbar \omega_{\alpha}}} \left[\hat{p}_{\alpha} + i\omega_{\alpha} \left(\hat{q}_{\alpha} - \frac{\boldsymbol{\lambda}_{\alpha}\cdot \hat{\boldsymbol{\mu}}}{\omega_{\alpha}}\right) \right] . \label{eq:create-length}
\end{align}
For a more detailed discussion on the definition of photonic observables, we refer the reader to App.~\ref{sec:observables}, where we present observables as defined in length and velocity (momentum) gauges commonly used in strongly coupled light-matter systems~\cite{schaefer2020}. We perform numerical exact diagonalizations of the model Hamiltonian of the atom-photon system in the length gauge. In Fig.~(\ref{fig:pt-occup-per-mode}) we show the mean photon occupation for the discretized continuum sampled for different light-matter coupling strengths $\lambda$. We find that the lower-lying photon modes have a higher photon occupation, since they interact more strongly with the atomic system. This is due to the fact that if we assume that all modes in the continuum have the same coupling strength $\lambda$, then the effective coupling strength behaves approximately as $\lambda/\sqrt{\omega_{\alpha}}$. These results highlight that the influence of the electromagnetic continuum on matter ground-state properties saturates with increasing photon mode energies. We note that the same trend of photon occupation per mode occurs in the molecular light-matter system. We further see that the stronger the coupling $\lambda$, the higher the photon occupation. The results of Fig.~(\ref{fig:pt-occup-per-mode}), which are relatively generic for coupled light-matter systems in the long-wavelength approximation, already point toward an approximation strategy, where only specific modes are treated and the rest subsumed or effectively discarded. However, we note that the influence of the higher-lying modes also depends on the dimensionality of the problem. In two and three dimensions, while they effectively have a smaller interaction strength due to $\lambda/\sqrt{\omega_{\alpha}}$, there are many more modes within a certain mode energy range. In fact, the necessity of a corresponding ultraviolet cut-off as well as a subsequent mass renormalization procedure in the long-wavelength approximation becomes specifically apparent in those cases with higher dimensions~\cite{welakuh2025}.

\begin{figure}
\includegraphics[width=1.0\columnwidth]{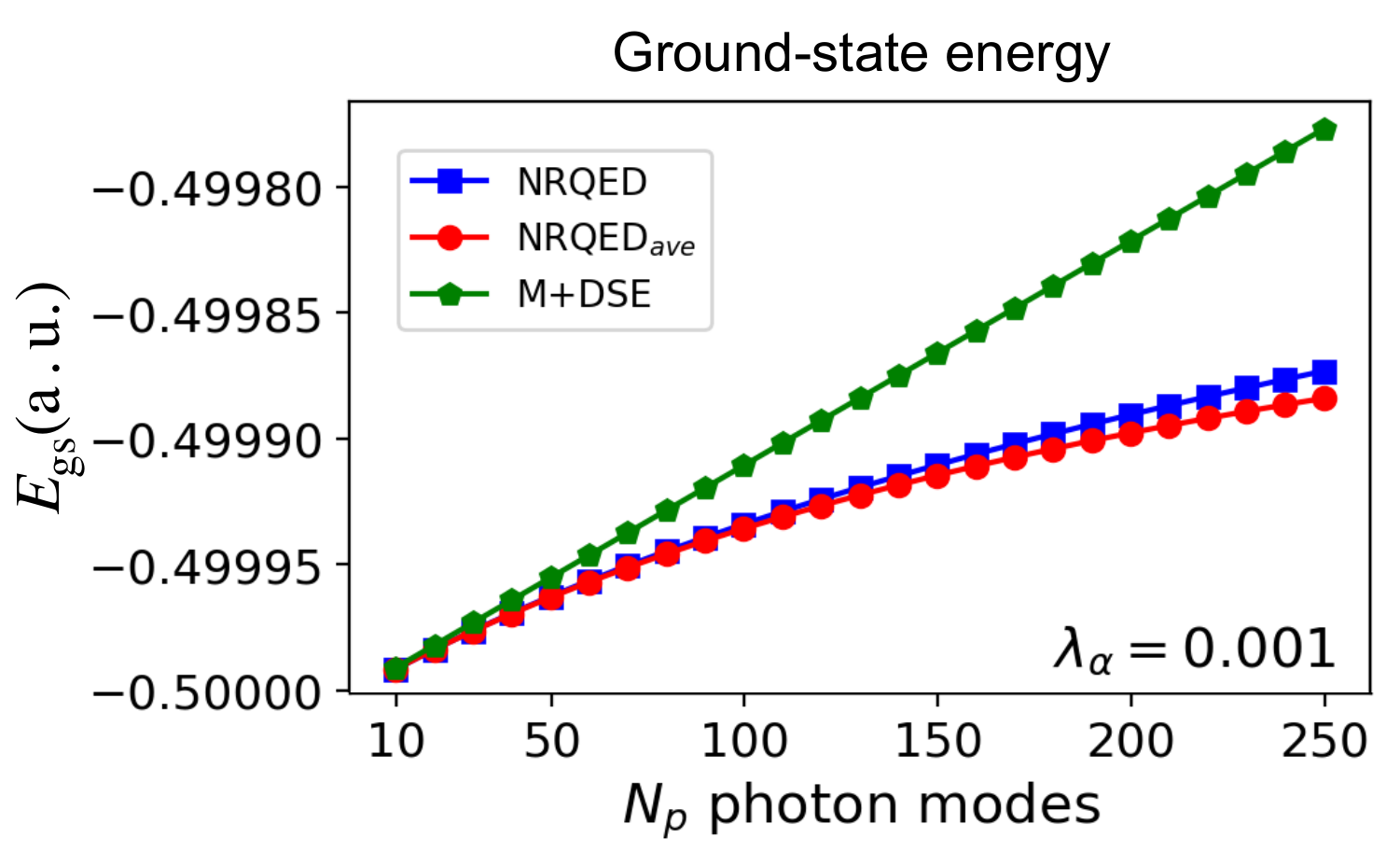}
\caption{Comparison of the correlated atomic ground-state energy for increasing number of modes showing a qualitative agreement between the result of NRQED and ``NRQED$_\text{ave}$'' while (M$+$DSE) deviates strongly for higher photon modes. }
\label{fig:gs-energy-NRQED-atom}
\end{figure}

Now, we investigate how the approximation strategies discussed above compare to the non-perturbative exact solutions of the coupled light-matter system. We show in Fig.~(\ref{fig:gs-energy-NRQED-atom}a) how the approximation strategy  ``NRQED$_\text{ave}$'' qualitatively approximates the non-perturbative NRQED results while the matter degrees including the dipole-self energy (M$+$DSE) deviates strongly especially for high-lying photon modes (i.e. $N_{p}>100$). The approximation strategy (M$+$DSE) deviates strongly from the NRQED result, since we can include an arbitrarily large number of photon modes, making the dipole self-energy a dominant contribution of Eq.~(\ref{eq:el-pt-hamiltonian-length-DSE}). 

Let us comment on the strong deviation of (M$+$DSE) from NRQED. We observe that (M$+$DSE) exhibits a clear linear trend as a function of number of photon modes $N_p$. On the other hand, the exact calculations coming from NRQED and the approximate NRQED$_{\rm{ave}}$ exhibit non-linear (approximately $\sqrt{N_p}$) dependence on the amount of photon modes $N_p$. The fundamental difference of (M$+$DSE) from the other two methods is that it totally neglects the photonic Hilbert space associated with the operators $\{\hat{q}_{\alpha},\hat{p}_{\alpha}\}$ and only treats the static contribution of the DSE. The NRQED$_{\rm{ave}}$ approximation includes the photonic Hilbert space and light-matter correlation, and as a consequence is in much better agreement with exact results of NRQED. Thus, we can understand the failure of the (M$+$DSE) as a failure of a mean-field approach which does not include light-matter correlation. On the other hand, the performance of NRQED$_{\rm{ave}}$ approximation is quite remarkable as it manages with just one effective average mode to capture qualitatively the exact multimode behavior. 

A physical interpretation of the result in Fig.~(\ref{fig:gs-energy-NRQED-atom}) is that the ground-state energy deviates from the uncoupled atomic value (i.e. $E_{0}=-0.5$~a.u.), indicating that the coupled system becomes increasingly correlated as the number of photon modes grows. Such enhanced correlations manifest as a reinforcement of particle confinement within the atomic binding potential. We do not show the results for NRQED$_{\rm{low}}$ in Fig.~(\ref{fig:gs-energy-NRQED-atom}) since we considered only a single photon mode which is the lowest frequency mode that interacts strongly with the atomic system while the comparison with the other approximation strategies considers $N_{p} \geq 10$ photon modes. Besides, NRQED$_{\rm{low}}$ corresponds to the exact NRQED for $N_p=1$.

\begin{figure}[bth]
\includegraphics[width=1.0\columnwidth]{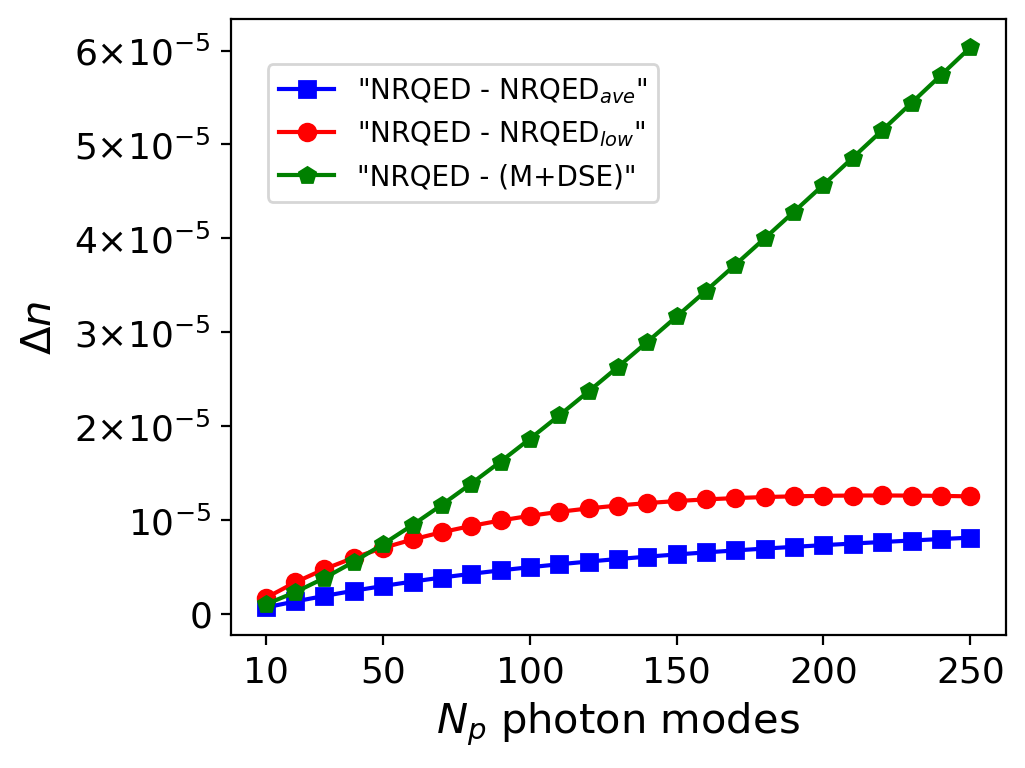}
\caption{The integrated atomic ground-state electron density difference between the NRQED result and the different approximations strategies, namely NRQED$_{\rm ave}$ (blue), NRQED$_{\rm low}$ (red), and $\rm (M+DSE)$ (green). We observe that $\rm (M+DSE)$ strongly deviates from the exact result for $N_p>50$, while NRQED$_{\rm low}$ has better performance but still performs worst than NRQED$_\text{ave}$. Thus, NRQED$_\text{ave}$ has the best performance when compared to the other methods.}
\label{fig:gs-density-diff}
\end{figure}

It is important to investigate how the different approximation strategies perform when we look at different observables of the light-matter coupled system. To do this, we compute the integrated ground-state electron density difference which is defined for a one-dimensional system as
\begin{align}
\Delta n = \int \! dx \, |n(x) - n'(x)| \, . \label{eq:gs-density-diff}
\end{align}
Here, $n(x)$ and $n'(x)$ represents the electron densities resulting from the exact Pauli-Fierz description (reference density) and an approximate description of a coupled light-matter system, respectively. For this quantity, the approximation is considered to perform well when $\Delta n \rightarrow 0$, that is, when the density difference between the non-perturbative exact NRQED result and any approximation strategy is very small. 

\begin{figure}
\includegraphics[width=1.0\columnwidth]{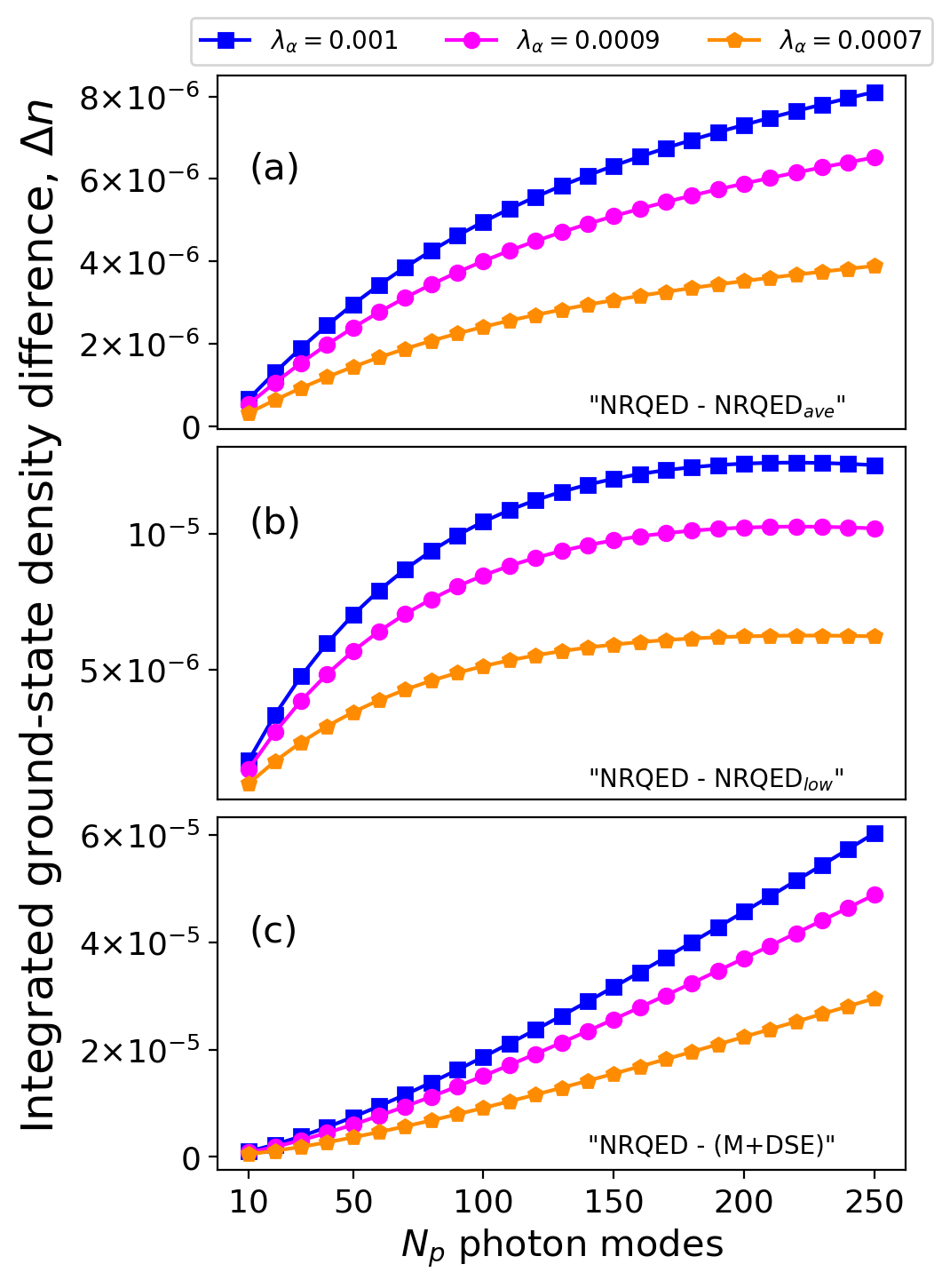}
\caption{The integrated atomic ground-state electron density difference for different light-matter coupling strengths for (a) ``NRQED$-$NRQED$_\text{ave}$'', (b) ``NRQED$-$NRQED$_\text{low}$'' and (c) ``NRQED$-$(M$+$DSE)'' where the density difference increases with $\lambda_{\alpha}$. We observe that NRQED$_\text{ave}$ performs the best as the deviations from the exact NRQED are an order of magnitude smaller ($\sim 10^{-6}$) in comparison to the other methods ($\sim 10^{-5}$).  }
\label{fig:gs-density-diff-lambda}
\end{figure}

In Fig.~(\ref{fig:gs-density-diff}) we show how the quantity $\Delta n$ between NRQED and ``NRQED$_\text{ave}$'' (denoted ``NRQED$-$NRQED$_\text{ave}$'') compare to ``NRQED$-$NRQED$_\text{low}$'', and ``NRQED$-$(M$+$DSE)''. We find that ``NRQED$_\text{ave}$'' has the best performance since its $\Delta n < 10^{-5}$. The deviation of the other two methods exceed $10^{-5}$, and particularly the deviation of (M$+$DSE) approximation increases linearly as function of $N_p$. The NRQED$_\text{low}$ method has a converging trend as function of $N_p$ but it is outperformed by NRQED$_\text{ave}$. Furthermore, in Fig.~(\ref{fig:gs-density-diff-lambda}) we show for the different approximation strategies how the quantity $\Delta n$ performs when we vary the light-matter coupling strength. We find the same trend for all approximation strategies where the different approximations perform well for weaker light-matter coupling strengths as $\Delta n \rightarrow 0$ for smaller values of $\lambda$.

\subsection{Molecule-photon system}
\label{subsubsec:molecular-system}

We now consider how the different approximation strategies perform when applied to effectively describe a molecule interacting with the discretized electromagnetic continuum. Here, we consider a one-dimensional model of the hydrogen molecule H$_{2}$ (see App.~\ref{sup:1D-molecular-hydrogen} for details) coupled to the same discretized continuum of photon modes discussed above. For this system, we compute the classical ground-state dissociation energy $D_{e}$ non-perturbatively through numerical exact diagonalizations of the Hamiltonian describing the molecule-photon system and make a comparison to the results obtained when we employ the different approximation strategies discussed above. 

\begin{figure}[bth]
\includegraphics[width=1.0\columnwidth]{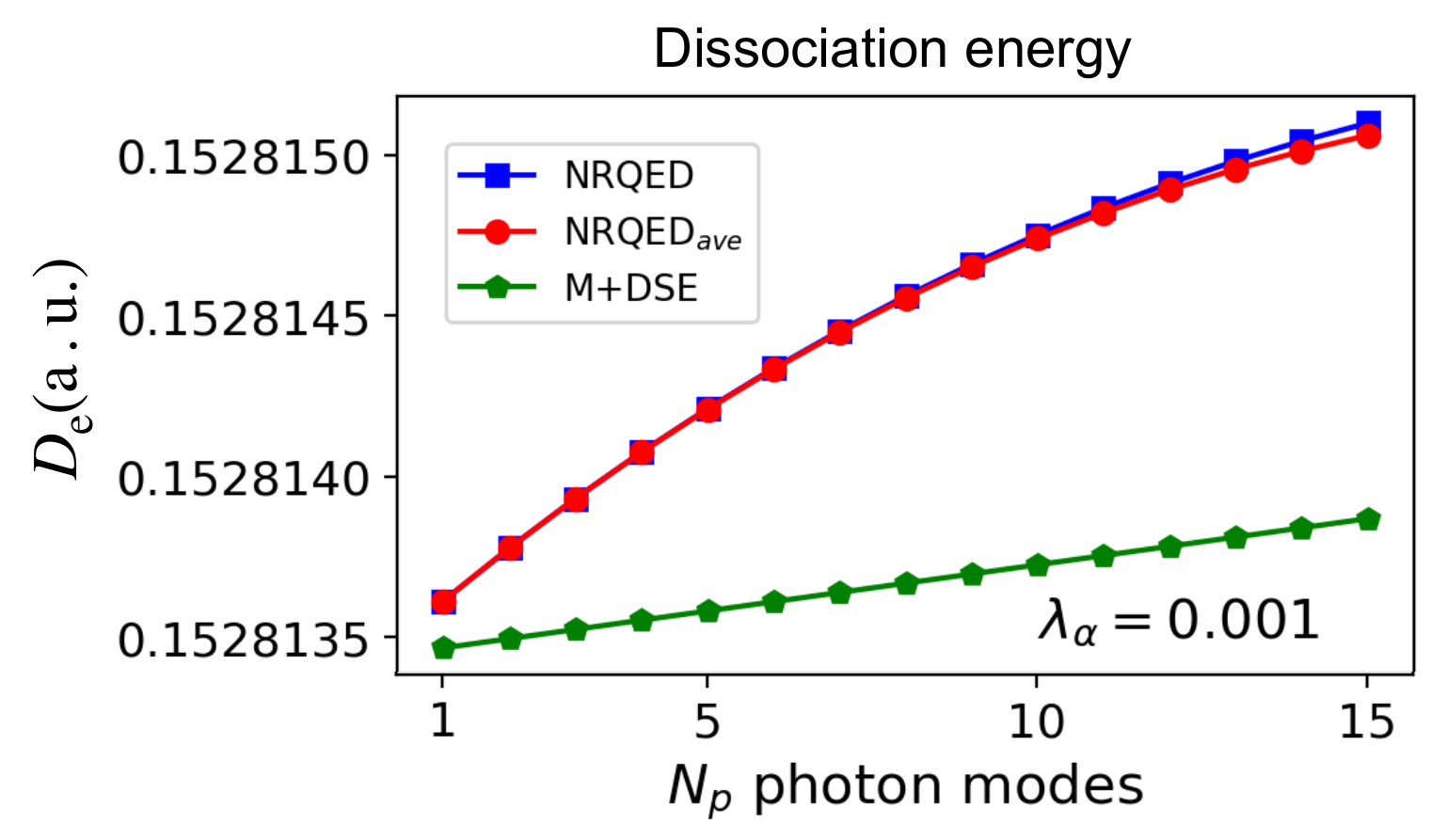}
\caption{Comparison of the dissociation energy of the ground-state of molecular H$_{2}$ for increasing number of photon modes where the (M$+$DSE) approximation strategy deviates from the exact NRQED result and the ``NRQED$_\text{ave}$'' qualitatively approximates the NRQED results. }
\label{fig:gs-PoPES}
\end{figure}

The results are shown in Fig.~(\ref{fig:gs-PoPES}a) where the ``NRQED$_\text{ave}$'' results are qualitatively and quantitatively very close to the exact NRQED result, while the (M$+$DSE) deviates strongly. As in the case of the atom-photon system, we attribute the failure of the (M$+$DSE) approximation due to the lack of light-matter correlations. Thus, similarly to the case of the atomic system we find again that NRQED$_\text{ave}$ is the best approximation strategy. In Fig.~(\ref{fig:gs-PoPES}b), we show the photon-mode-dependent dissociation energy of the ground-state PES for different light-matter couplings where we find that it becomes more difficult to break a chemical bond when the system is strongly coupled. We note that for the molecular dissociation energy we do not show separately the NRQED$_{\rm low}$ as it corresponds to exact NRQED for $N_p=1$.

\subsection{Two-dimensional quantum ring coupled to photons}
\label{subsubsec:molecular-system}

Lastly, we demonstrate the versatility of the approximation strategies discussed above by considering a higher-dimensional matter system coupled to the discretized electromagnetic continuum. We consider a model of a semiconductor GaAs quantum ring studied extensively in previous works~\cite{rasanen2007,flick2015,flick2017b,welakuh2021,welakuh2022c} where a single effective electron is restricted to two-dimensional real-space ($\hat{\textbf{r}} = \hat{x} \textbf{e}_{x} + \hat{y} \textbf{e}_{y}$) and trapped in a Mexican hat potential (see App.~\ref{sup:quantum-ring} for details). For the cavity setup, we consider the two independent polarization directions of the cavity modes such that the coupling is given as $\boldsymbol{\lambda}_{\alpha} = \lambda_{\alpha,x} \, \textbf{e}_{x} +  \lambda_{\alpha,y} \, \textbf{e}_{y}$, where $\lambda_{\alpha,x}$ and $\lambda_{\alpha,y}$ are the coupling strengths in the $x$- and $y$-directions, respectively. We include only the 30 lowest frequency photon modes for each of the polarization directions and we choose the couplings such that $\lambda_{\alpha,x} > \lambda_{\alpha,y}$ which breaks the intrinsic inversion and rotation symmetry of the quantum ring model~\cite{welakuh2022c}. Our choice of using only the 30 lowest frequency photon modes is motivated by the results in Fig.~(\ref{fig:pt-occup-per-mode}) where we see that these modes have the largest effect on the properties of the matter system. In addition, by including this limited amount of photonic modes allows to accurately compute the exact spectrum of the coupled matter-photon system. 

\begin{figure}[bth]
\includegraphics[width=1.0\columnwidth]{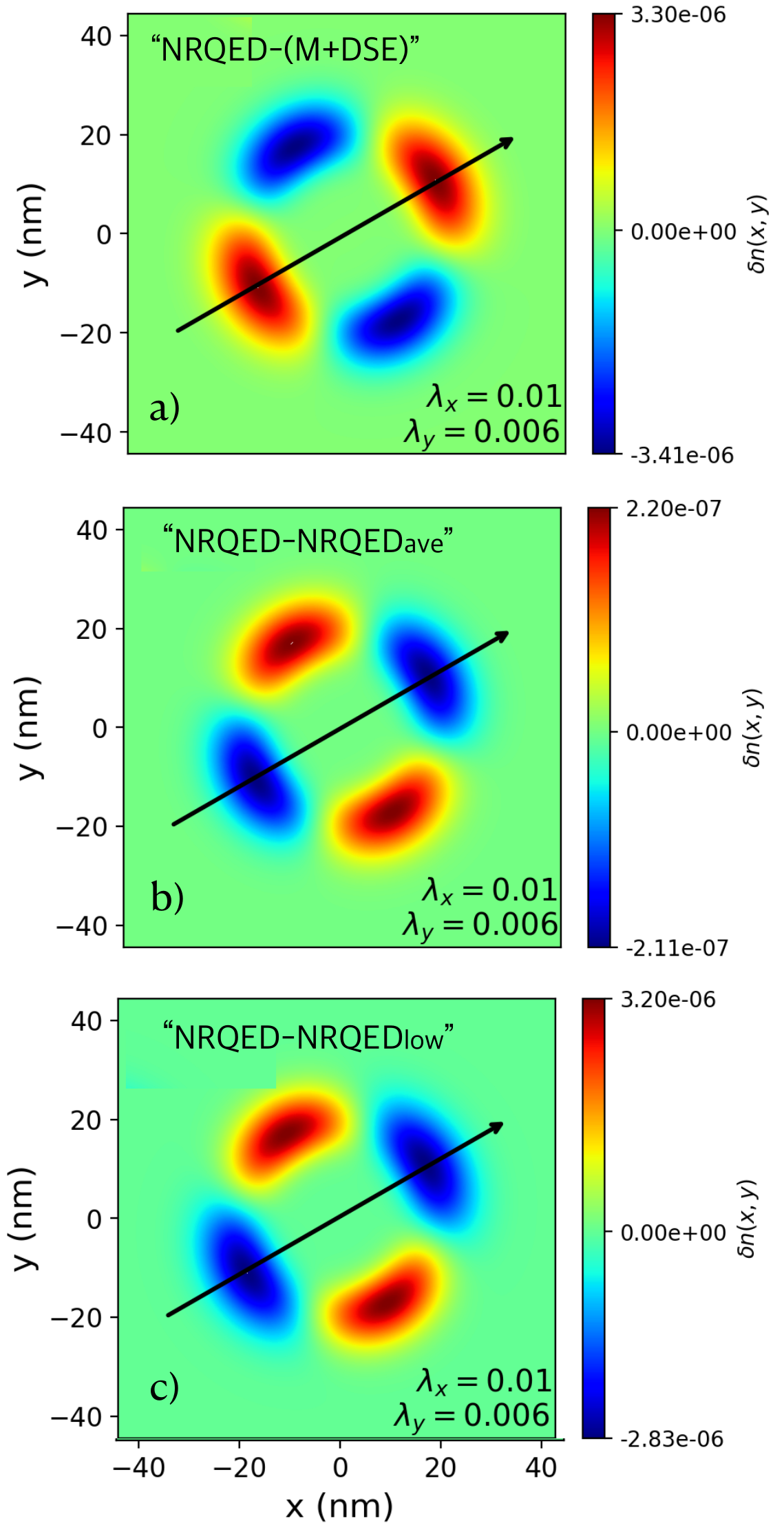}
\caption{The ground-state density difference between the exact NRQED and the the three approximate methods for the GaAs quantum ring for $\lambda_{x}=0.01$ and $\lambda_{y}=0.006$. In (a) we show density difference of M+DSE method, in (b) the density of difference of NRQED$_{\rm ave}$, and (c) density for NRQED$_{\rm low}$. We find that NRQED$\rm{ave}$ is the most accurate method showing deviations from the exact results of the order $10^{-7}$, while the other two methods deviate at the order $10^{-6}$. }
\label{ring}
\end{figure}

We determine the exact electronic ground-state density of the quantum ring coupled to the photonic using the exact NRQED and compare it to the three approximation methods, namely NRQED$_{\rm ave}$, NRQED$_{\rm low}$, and M+DSE. In Fig.~\ref{ring} we show the result for fixed couplings in the $x$ and the $y$ directions, given by $\lambda_x=0.01$ and $\lambda_y=0.006$ respectively. By comparing the density deviations for the different methods shown in panels (a)--(c) of Fig.~\ref{ring} we observe that the smallest deviations appear for the NRQED$_{\rm ave}$ method. For the methods (M+DSE) and NRQED$_{\rm low}$ in panels (a) and (c) the deviations are of the order $10^{-6}$, while for NRQED$_{\rm ave}$ the deviations are one order smaller reaching $10^{-7}$. Thus, we find that also for the two-dimensional quantum ring model the NRQED$_{\rm ave}$ approximation method outperforms the other two methods. This is a very important finding because it allows us to conclude that for all examined models-atomic, molecular and two dimensional quantum ring-the averaging approximation NRQED$_{\rm ave}$ is the best approximation method for describing ground-state properties of the multi-mode light-matter coupled system. 

Further, we note that the (M+DSE) method generates deviations from the exact NRQED results in the \emph{opposite} directions when compared to the other two approximation methods. This is a consequence of the fact that (M+DSE) incorporates only the scalar potential of the dipole self-energy without the inclusion of light-matter correlations. Lastly, it is important to highlight that for completeness in Appendix~\ref{sup:quantum-ring} we provide the exact NRQED density and the approximated density for each method for several different values of the light-matter couplings.

\section{Summary and Outlook}
\label{sec:conclusion-outlook}

In this work, we presented three different approximation strategies that can be employed for the description of equilibrium properties of matter strongly coupled to a multi-mode photon field. The first strategy considers matter coupling to the transverse electromagnetic field only through the self-polarization of matter due to the photon field, also known as dipole self-energy in the length gauge. The second and third approaches retain all contributions of a coupled light-matter system, but one approximation strategy includes only the highest occupied photon mode, while the other averages over the photonic modes to obtain an effective single photon mode. We applied the approximation methods to different model systems, including a one-dimensional atom and molecule, and a two-dimensional quantum ring. We compared the results of the approximate methods to the numerically exact solutions of the coupled light-matter systems. For all tested model systems, we find that the averaging approximation method performs the best and describes most accurately the ground-state properties of the coupled system. We attribute the effectiveness of the averaging approximation to its ability to balance the dominant contributions from the lowest photonic modes with the corrective influence of higher-frequency modes, achieved through averaging over the entire photonic spectrum. On the other hand, the lowest mode approximation (highest occupied mode) neglects completely all the high frequency modes and thus misses important contributions. The dipole self-energy approximation is less accurate because it misses the explicit light-matter correlations. It is effectively a mean-field approximation that mostly works for weak coupling and for small number of photon modes. 

The importance of this averaging approximation method is that it can circumvent the high computational cost due to explicitly including the full photonic continuum of modes normally encountered with first-principles theoretical approaches capable of treating the coupled light-matter~\cite{ruggenthaler2014,flick2015,flick2019,welakuh2022,haugland2020,flick2018}. We note that beyond the simple (discretized) continua investigated in this work, for realistic cavities a similar averaging procedure for ground-state properties has been proposed in Ref.~\cite{svendsen2023theory}. Here we can substantiate these theoretical considerations by concrete examples and show that effective-mode approximations can capture ground-state modifications of realistic quantum systems. Thus, this work suggests an efficient and low-cost computational method for the treatment of quantum matter strongly coupled to multi-mode quantized electromagnetic fields, which can be applied to a wide range of systems, enabling the first-principles simulation of quantum systems in realistic photonic environments.


\section*{Acknowledgment}

This work was funded by the European Union under the European Research Council (ERC-2024-SyG- 101167294 ; UnMySt) and the Simons Foundation (Grant 839534, MET). We acknowledge support from the Max Planck-New York City Center for Non-Equilibrium Quantum Phenomena. The Flatiron Institute is a division of the Simons Foundation. We also acknowledge support of the Cluster of Excellence “CUI: Advanced Imaging of Matter” of the Deutsche Forschungsgemeinschaft (DFG), EXC 2056, project ID 390715994 and the Grupos Consolidados (IT1453-22). 


\vspace{10em}

\bibliography{01_light_matter_coupling}


\appendix

\section{Observables in the velocity and length gauges}
\label{sec:observables}

In this section, we present observables as defined in the length and velocity gauge descriptions of a coupled light-matter system. The purpose here is to stress that care needs to be taken when using observables normally defined in the velocity gauge when working with the length gauge Hamiltonian. We start with the velocity gauge Hamiltonian given by
\begin{align}
\hat{H}_{\text{V}} &= \sum\limits_{l=1}^{N_{e}}\frac{1}{2m} \left(\hat{\textbf{p}}_{l} - |e| \hat{\textbf{A}}_{\rm V}\right)^{2} + \frac{1}{2} \sum\limits_{l\neq j}^{N_{e}} w(|\hat{\textbf{r}}_{l} - \hat{\textbf{r}}_{j}|) \nonumber \\
& \quad + \sum\limits_{l=1}^{N_{n}}\!\frac{1}{2M_{l}} \! \left(\!\hat{\textbf{P}}_{l} + Z_{l}|e|\hat{\textbf{A}}_{\rm V}\!\right)^{2} \! - \sum\limits_{l=1}^{N_{e}} \! \sum\limits_{j=1}^{N_{n}} Z_{j}w(|\hat{\textbf{r}}_{l} - \hat{\textbf{R}}_{j}|)   \nonumber \\
& \quad + \frac{1}{2} \sum\limits_{l\neq j}^{N_{n}} Z_{l}Z_{j}w(|\hat{\textbf{R}}_{l} - \hat{\textbf{R}}_{j}|) + \sum_{\alpha=1}^{N_{p}} \hbar \omega_{\alpha} \left(\hat{a}^{\dagger}_{\alpha, \rm V} \hat{a}_{\alpha, \rm V}+\tfrac{1}{2}\right)  .\label{eq:velocity-gauge-hamiltonian}
\end{align}
Here, the energy of the quantized electromagnetic field is given in terms of bosonic creation $\hat{a}^{\dagger}_{\alpha, \rm V}$ and annihilation $\hat{a}_{\alpha, \rm V}$ operators that satisfy the commutation relations
\begin{align}
\left[\hat{a}_{\alpha, \rm V}, \hat{a}^{\dagger}_{\alpha', \rm V}\right]= \delta_{\alpha,\alpha', } \, , \quad \left[\hat{a}_{\alpha, \rm V}, \hat{a}_{\alpha', \rm V}\right]= \left[\hat{a}_{\alpha, \rm V}^{\dagger}, \hat{a}_{\alpha'}^{\dagger}\right] = 0 \, .
\end{align}
We note here that the index $V$ attached to operators is used to indicate that we are in the velocity gauge. When the index $L$ is used later, this indicates that we are working with operators in the length gauge. The quantized vector potential in Eq.~(\ref{eq:velocity-gauge-hamiltonian}) is given in terms of the bosonic operators as
\begin{align}
\hat{\textbf{A}}_{\rm V} = \sum\limits_{\alpha=1}^{N_{p}} \,\boldsymbol{\lambda}_{\alpha} \sqrt{ \frac{\hbar}{2\omega_{\alpha}}}\left(\hat{a}_{\alpha, \rm V} + \hat{a}_{\alpha, \rm V}^{\dagger}\right) = \sum\limits_{\alpha=1}^{N_{p}} \,\boldsymbol{\lambda}_{\alpha} \, \hat{q}_{\alpha, \rm V} \,   . \label{eq:vector-potential}
\end{align}
Here, we used the definition of the bosonic creation and annihilation operators in terms of the photon coordinate and conjugate momentum~\cite{gerry2005} which are given respectively by
\begin{equation}
\begin{aligned}
\hat{a}_{\alpha, \rm V}^{\dagger} &= \frac{1}{\sqrt{2\hbar \omega_{\alpha}}} \left(\omega_{\alpha} \hat{q}_{\alpha, \rm V} - i \hat{p}_{\alpha, \rm V} \right) , \\
\hat{a}_{\alpha, \rm V} &= \frac{1}{\sqrt{2\hbar \omega_{\alpha}}} \left(\omega_{\alpha} \hat{q}_{\alpha, \rm V} + i \hat{p}_{\alpha, \rm V} \right) . 
\end{aligned} \label{eq:a-adag-VG}
\end{equation}
From Eq.~(\ref{eq:a-adag-VG}), the photon coordinate and its conjugate momentum can be expressed in terms of the bosonic creation and annihilation operators as follows
\begin{equation}
\begin{aligned}
\hat{q}_{\alpha, \rm V} &= \sqrt{ \frac{\hbar}{2 \omega_{\alpha}}} \left(\hat{a}_{\alpha, \rm V} + \hat{a}^{\dagger}_{\alpha, \rm V} \right) , \\
\hat{p}_{\alpha, \rm V} &= -i\sqrt{ \frac{\hbar \omega_{\alpha}}{2 }} \left(\hat{a}_{\alpha, \rm V} - \hat{a}^{\dagger}_{\alpha, \rm V} \right) . 
\end{aligned} \label{eq:pt-coord-momen-VG}
\end{equation}
The photon coordinate and its conjugate momentum satisfy the commutation relations 
\begin{align}
\bigl[\hat{q}_{\alpha, \rm V}, \hat{p}_{\alpha', \rm V} \bigr] = i\hbar \delta_{\alpha,\alpha'} \, \quad \bigl[\hat{q}_{\alpha, \rm V}, \hat{q}_{\alpha', \rm V} \bigr] = \bigl[\hat{p}_{\alpha, \rm V}, \hat{p}_{\alpha', \rm V} \bigr] = 0 \, . \label{eq:commutation-p-q-VG}
\end{align}
It is common for finite systems to study the coupled light-matter system in a unitarily equivalent form. This is done by first transforming Eq.~(\ref{eq:velocity-gauge-hamiltonian}) according to~\cite{rokaj2017}
\begin{align}
\hat{H}_{\text{L}}' = \hat{U}^{\dagger} \hat{H}_{\text{V}} \hat{U} \quad \textrm{where} \quad  \hat{U} = \exp \left(\frac{i}{\hbar} \hat{\textbf{A}}_{\rm V}\cdot\hat{\boldsymbol{\mu}}\right) \, , \label{eq:velocity-to-length-gauge}
\end{align}
where the total dipole operator is given by
\begin{align}
\hat{\boldsymbol{\mu}} = -\sum\limits_{l=1}^{N_{e}} |e| \, \hat{\textbf{r}}_{l} +  \sum\limits_{l=1}^{N_{n}}Z_{l}|e|\hat{\textbf{R}}_{l} \, . \label{eq:dipole}
\end{align}
In a next step, a canonical transformation is performed that swaps the canonical photon coordinates and momenta~\cite{tokatly2013} as follows
\begin{align}
\hat{p}_{\alpha,\rm L} \longrightarrow - \omega_{\alpha} \hat{q}_{\alpha, \rm L} \quad \textrm{and} \quad \hat{q}_{\alpha, \rm L}  \longrightarrow \frac{1}{\omega_{\alpha}} \hat{p}_{\alpha, \rm L} \, ,\label{eq:transform-coord-momen}
\end{align}
while maintaining the commutation relations between the displacement coordinate $\hat{q}_{\alpha, \rm L}$ and $\hat{p}_{\alpha, \rm L}$, its conjugate momentum as in Eq.~(\ref{eq:commutation-p-q-VG}). The resulting Hamiltonian is the length form of the Pauli-Fierz Hamiltonian given in Eq.~(\ref{eq:length-gauge-hamiltonian}). 

As we transformed the velocity gauge Hamiltonian to the unitarily equivalent length gauge Hamiltonian using Eqs.~(\ref{eq:velocity-to-length-gauge}) and (\ref{eq:transform-coord-momen}), we \textit{must} perform the same operation to any operator defined in the velocity gauge to get its unitary equivalent form in the length gauge. That is, for an operator $\hat{O}_{\rm V}$ defined in the velocity gauge, its unitarily equivalent length gauge form is obtained using the relation $\hat{O}_{\text{L}}' = \hat{U}^{\dagger} \hat{O}_{\text{V}} \hat{U}$ together with Eq.~(\ref{eq:transform-coord-momen}). As an example, we can also define new creation and annihilation operators for the length gauge, but we note that they are \textit{not} the original photonic operators as defined in Eq.~(\ref{eq:a-adag-VG}). Instead, they are mixed light-matter objects and the original annihilation and creation operators now represented in length gauge are~\cite{rokaj2017,schaefer2020}
\begin{align}
\hat{a}_{\alpha,L} 
&= \frac{1}{\sqrt{2\hbar \omega_{\alpha}}} \left[\hat{p}_{\alpha, \rm L} - i\omega_{\alpha} \left(\hat{q}_{\alpha, \rm L} - \frac{\boldsymbol{\lambda}_{\alpha}\cdot\hat{\boldsymbol{\mu}}}{\omega_{\alpha}}\right) \right] , \label{eq:annahil-length} \\
\hat{a}_{\alpha,L}^{\dagger} 
&= \frac{1}{\sqrt{2\hbar \omega_{\alpha}}} \left[\hat{p}_{\alpha, \rm L} + i\omega_{\alpha} \left(\hat{q}_{\alpha, \rm L} - \frac{\boldsymbol{\lambda}_{\alpha}\cdot\hat{\boldsymbol{\mu}}}{\omega_{\alpha}}\right) \right] . \label{eq:create-length}
\end{align}
From Eqs.~(\ref{eq:annahil-length}) and (\ref{eq:create-length}), we can verify that $\hat{q}_{\alpha,\text{L}}$ and  $\hat{p}_{\alpha,\text{L}}$ can be expressed in terms of  $\hat{a}_{\alpha,\text{L}}$ and $\hat{a}^{\dagger}_{\alpha,\text{L}}$ after performing the transformation in Eq.~(\ref{eq:transform-coord-momen}) such that we find
\begin{equation}
\begin{aligned}
\hat{q}_{\alpha, \rm L} &= \sqrt{ \frac{\hbar}{2 \omega_{\alpha}}} \left(\hat{a}_{\alpha, \rm L} + \hat{a}^{\dagger}_{\alpha, \rm L} \right) , \\
\hat{p}_{\alpha, \rm L} &= -i\sqrt{ \frac{\hbar \omega_{\alpha}}{2 }} \left(\hat{a}_{\alpha, \rm L} - \hat{a}^{\dagger}_{\alpha, \rm L} \right) . 
\end{aligned} \label{eq:pt-coord-momen-LG}
\end{equation}


We now present different observables in the velocity gauge and their corresponding definitions in the length gauge and demonstrate how the different representations of the observables lead to the same results. The outcome of these results stresses the importance of performing the transformation of Eq.~(\ref{eq:velocity-to-length-gauge}) together with Eq.~(\ref{eq:transform-coord-momen}) to obtain the correct expression of observables in the length gauge Eq.~(\ref{eq:length-gauge-hamiltonian}). We will mainly focus on photonic observables since they are defined in terms of $\hat{a}_{\alpha,\rm V}$ and $\hat{a}^{\dagger}_{\alpha,\rm V}$ in the velocity gauge, but have different expressions in the length gauge as given in Eqs.~(\ref{eq:annahil-length}) and (\ref{eq:create-length}). Note that the observables presented below are considered under the dipole approximation~\cite{tannoudji1989}.
\\

\noindent \textbf{The transverse electric-field}: This is the electric field component of the electromagnetic field that is quantized and lies transverse (perpendicular) to the direction of wave propagation. The transverse electric field in the velocity and length gauge is defined as
\begin{equation}
\begin{aligned}
\hat{\textbf{E}}_{\text{V}} &= \sum_{\alpha=1}^{N_{p}}\,i \, \boldsymbol{\lambda}_{\alpha} \sqrt{\frac{\hbar\omega_{\alpha}}{2}}\left(\hat{a}_{\alpha,\rm V} - \hat{a}_{\alpha,\rm V}^{\dagger}\right) = - \sum_{\alpha=1}^{N_{p}}\, \boldsymbol{\lambda}_{\alpha} \, \hat{p}_{\alpha,\rm V} \, , \\
\hat{\textbf{E}}_{\text{L}} &= \sum_{\alpha=1}^{N_{p}} \boldsymbol{\lambda}_{\alpha}\omega_{\alpha}   \left( \hat{q}_{\alpha,\rm L} -  \frac{\boldsymbol{\lambda}_{\alpha} \cdot\hat{\boldsymbol{\mu}}}{\omega_{\alpha}} \right) . \label{eq:E-field}
\end{aligned}
\end{equation}

\noindent \textbf{The squared-electric field}: It is the square of the transverse electric field operator that is central in calculating observables such as energy density, radiation pressure, or vacuum fluctuations~\cite{gerry2005}. The squared-electric field of the mode $\alpha$ in the velocity and length gauge is defined as
\begin{equation}
\begin{aligned}
\hat{\textbf{E}}_{\alpha,\text{V}}^{2} &= \boldsymbol{\lambda}_{\alpha}^{2} \left( \frac{\hbar\omega_{\alpha}}{2} \right) \left[2\hat{a}_{\alpha,\rm V}^{\dagger}\hat{a}_{\alpha,\rm V} + 1 - \hat{a}_{\alpha,\rm V}^{\dagger \, 2} - \hat{a}_{\alpha,\rm V}^{2} \right] =\boldsymbol{\lambda}_{\alpha}^{2} \, \hat{p}_{\alpha,\rm V}^{2} \, , \\
\hat{\textbf{E}}_{\alpha,\text{L}}^{2} &= \boldsymbol{\lambda}_{\alpha}^{2} \left[\omega_{\alpha}^{2}\hat{q}_{\alpha,\rm L}^{2} + \left(\boldsymbol{\lambda}_{\alpha} \cdot \hat{\boldsymbol{\mu}}  \right)^2 - 2\omega_{\alpha}\hat{q}_{\alpha,\rm L}\left(\boldsymbol{\lambda}_{\alpha} \cdot \hat{\boldsymbol{\mu}} \right)\right] . \label{eq:E-field-square}
\end{aligned}
\end{equation}

\noindent \textbf{The photon number operator}: It is a quantum mechanical operator that gives the number of photons with energy $\hbar\omega_{\alpha}$ in a specific quantum state or mode of the electromagnetic field~\cite{loudon2000,gerry2005}. The quantity is defined in the velocity and length gauge by
\begin{equation}
\begin{aligned}
\hat{n}_{\alpha,\text{V}} &= \hat{a}_{\alpha,\rm V}^{\dagger}\hat{a}_{\alpha,\rm V} = \frac{1}{2\hbar\omega_{\alpha}}(\hat{p}_{\alpha,\text{V}}^{2}  + \omega_{\alpha}^{2} \hat{q}_{\alpha,\text{V}}^{2}) - \frac{1}{2} \, , \\
\hat{n}_{\alpha,\text{L}} &=  \frac{1}{2\hbar\omega_{\alpha}}\left[\hat{p}_{\alpha,\rm L}^{2} + \omega^2_{\alpha}\left( \hat{q}_{\alpha,\rm L} - \frac{\boldsymbol{\lambda}_{\alpha}}{\omega_{\alpha}} \cdot \hat{\boldsymbol{\mu}} \right)^2\right] - \frac{1}{2} \, . \label{eq:photon-occupation}
\end{aligned}
\end{equation}

\noindent \textbf{The Mandel Q parameter}: This quantity measures the deviation of the photon statistics from a Poisson distribution~\cite{mandel1979} which is defined as
\begin{equation}
\begin{aligned}
Q_{\alpha,\text{V}} &=  \frac{\langle \hat{a}_{\alpha,\rm V}^{\dagger}\hat{a}_{\alpha,\rm V}^{\dagger}\hat{a}_{\alpha,\rm V}\hat{a}_{\alpha,\rm V}\rangle - \langle \hat{a}_{\alpha,\rm V}^{\dagger}\hat{a}_{\alpha,\rm V}\rangle^{2} }{\langle \hat{a}_{\alpha,\rm V}^{\dagger}\hat{a}_{\alpha,\rm V}\rangle} \, , \\
Q_{\alpha,\text{L}} &=  \frac{\langle \hat{a}_{\alpha,\text{L}}^{\dagger}\hat{a}_{\alpha,\text{L}}^{\dagger}\hat{a}_{\alpha,\text{L}}\hat{a}_{\alpha,\text{L}}\rangle - \langle \hat{a}_{\alpha,\text{L}}^{\dagger}\hat{a}_{\alpha,\text{L}}\rangle^{2} }{\langle \hat{a}_{\alpha,\text{L}}^{\dagger}\hat{a}_{\alpha,\text{L}}\rangle} \, .  \label{eq:mandel-q}
\end{aligned}
\end{equation}
For values within the range $-1 < Q_{\alpha,\textrm{V,L}} < 0$ indicates sub-Poissonian statistics (nonclassical light field), for $Q_{\alpha,\text{V,L}} > 0$ indicates super-Poissonian statistics (chaotic light) and $Q_{\alpha,\text{V,L}} = 0$ indicates Poissonian statistics (coherent state field).


\noindent \textbf{The current operator}: The total current operator, which is a sum of the paramagnetic and diamagnetic contribution in the velocity gauge, is given by
\begin{equation}
\begin{aligned}
 \hat{\textbf{j}}_{\text{V}} &= \frac{e}{m}\sum_{l=1}^{N_{e}}\, \hat{\textbf{p}}_{l} - \frac{N_{e}e^{2}}{m} \hat{\textbf{A}}_{\rm V} \, , \\
\hat{\textbf{j}}_{\text{L}} &= \frac{e}{m}\sum_{l=1}^{N_{e}}\hat{\textbf{p}}_{l} \, . \label{eq:total-current}
\end{aligned}
\end{equation}
This can be verified using the mode-resolved equation of motion of the quantized vector potential in the velocity and the length setting which is given by
\begin{equation}
\begin{aligned}
\left(\frac{d^2}{dt^2} + \omega_{\alpha}^2 \right)\hat{\textbf{A}}_{\alpha, \rm V} &= \frac{e}{m}\boldsymbol{\lambda}_{\alpha} \, \boldsymbol{\lambda}_{\alpha} \cdot \sum\limits_{l=1}^{N_{e}} \left(\hat{\textbf{p}}_{l} - \sum_{\alpha=1}^{N_{p}}\hat{\textbf{A}}_{\alpha, \rm V} \right) \, , \\
\left(\frac{d^2}{dt^2} + \omega_{\alpha}^2 \right)\hat{\textbf{A}}_{\alpha, \rm L} &= \frac{e}{m} \boldsymbol{\lambda}_{\alpha} \boldsymbol{\lambda}_{\alpha} \cdot \sum\limits_{l=1}^{N_{e}}\hat{\textbf{p}}_{l} \, . 
\label{eq:EOM-of-vector-potential}
\end{aligned}
\end{equation}
Here, the quantized vector potential in the velocity and length gauge is given, respectively, as
\begin{equation}
\begin{aligned}
\hat{\textbf{A}}_{\rm V} &=  \sum\limits_{\alpha=1}^{N_{p}} \boldsymbol{\lambda}_{\alpha} \, \hat{q}_{\alpha, \rm V} \, , \\
\hat{\textbf{A}}_{\text{L}} &= \sum\limits_{\alpha=1}^{N_{p}} \frac{\boldsymbol{\lambda}_{\alpha}}{\omega_{\alpha}}  \, \hat{p}_{\alpha, \rm L}  \, . 
\label{eq:vector-potential-VG-LG}
\end{aligned}
\end{equation}
Using the relation between the electric field and the vector potential $\hat{\textbf{E}} = - \frac{d}{dt} \hat{\textbf{A}}$, we determine the mode-resolved equation of motion of the quantized electric field in the velocity and length setting which is given by
\begin{equation}
\begin{aligned}
\left(\frac{d^2}{dt^2} + \omega_{\alpha}^2 \right)\hat{\textbf{E}}_{\alpha, \rm V} &=  \frac{e}{m} \boldsymbol{\lambda}_{\alpha} \, \boldsymbol{\lambda}_{\alpha} \cdot \sum\limits_{l=1}^{N_{e}} \left(  \frac{d\hat{\textbf{p}}_{l}}{dt} -  \sum\limits_{\alpha=1}^{N_{p}} \hat{\textbf{E}}_{\alpha, \rm V} \right) \, , \\
\left(\frac{d^2}{dt^2} + \omega_{\alpha}^2 \right)\hat{\textbf{E}}_{\alpha, \rm L} &= - \frac{e}{m} \boldsymbol{\lambda}_{\alpha}\sum\limits_{l=1}^{N_{e}} \boldsymbol{\lambda}_{\alpha} \cdot \frac{d\hat{\textbf{p}}_{l}}{dt} \, . 
\label{eq:EOM-of-electric-field}
\end{aligned}
\end{equation}
where the first derivatives of the vector potential in both gauges are given by
\begin{equation}
\begin{aligned}
\frac{d}{dt} \hat{\textbf{A}}_{\rm V} &=  \sum\limits_{\alpha=1}^{N_{p}} \boldsymbol{\lambda}_{\alpha} \; \hat{p}_{\alpha, \rm V} \, , \\
\frac{d}{dt} \hat{\textbf{A}}_{\text{L}} &= -\boldsymbol{\lambda}_{\alpha} \, \omega_{\alpha}\left(\hat{q}_{\alpha, \rm L} \!-\! \frac{\boldsymbol{\lambda}_{\alpha}}{\omega_{\alpha}} \cdot \hat{\boldsymbol{\mu}}  \right)  \, . 
\label{eq:vector-potential-VG-LG-d-dt}
\end{aligned}
\end{equation}
Equation (\ref{eq:vector-potential-VG-LG-d-dt}) is related to eq.~(\ref{eq:E-field}) via the relation $\hat{\textbf{E}} = - \frac{d}{dt} \hat{\textbf{A}}$.\\

\begin{figure}[bth]
\includegraphics[width=1.0\columnwidth]{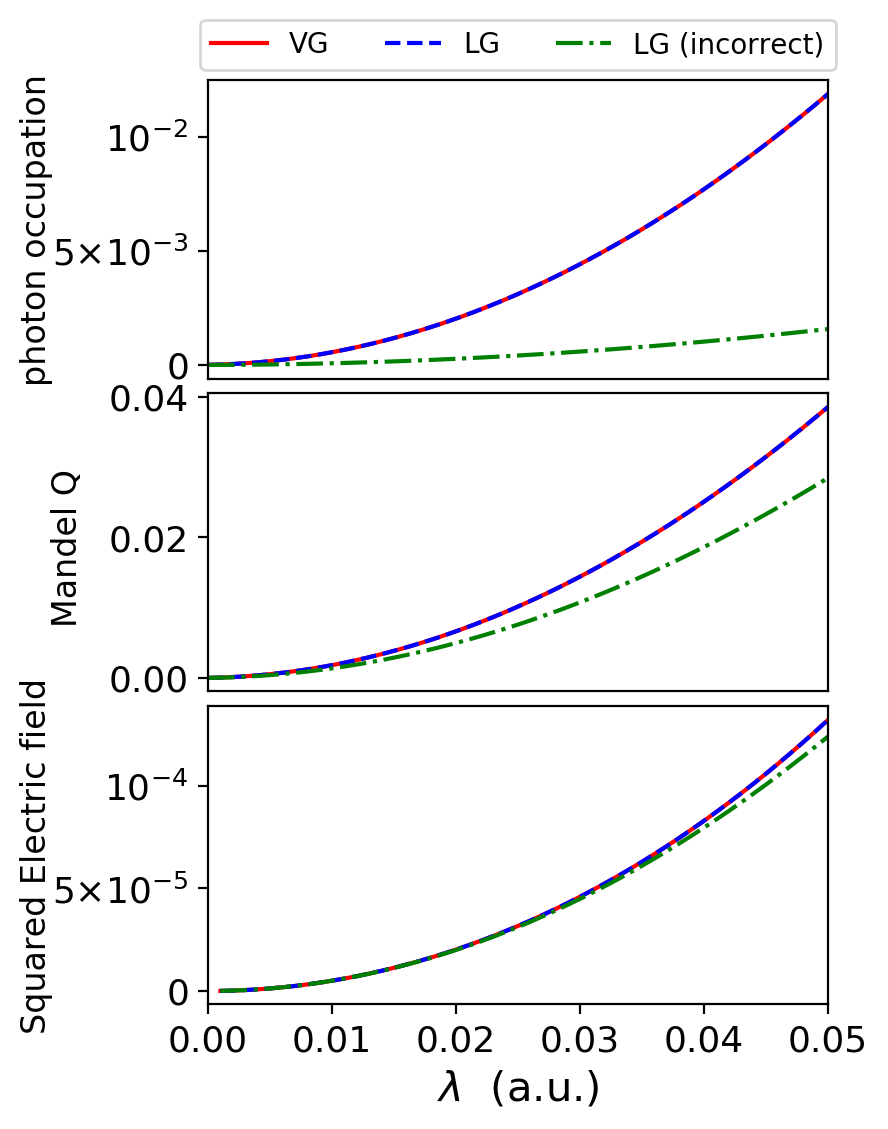}
\caption{Observables of the ground-state of the correlated atom-photon system, where the results of the velocity- and length-gauge numerical calculations agree for the range of the light-matter coupling strength. The incorrect length gauge results (see Eqs.~\ref{eq:mixed-observables} and \ref{eq:a-adag-LG-incorrect}) differ for increasing $\lambda$.}
\label{fig:gs-observables-LG-VG}
\end{figure}

\noindent We will now show how different observables defined in the respective length and velocity gauges. To exemplify this, we consider a one-dimensional model of atomic hydrogen (see App.~\ref{sup:1D-atomic-hydrogen} for details) coupled to a single-photon mode in both the length and velocity gauge descriptions. As observables, we compute the expectation values of the photon number operator (see Eq.~(\ref{eq:photon-occupation})), Mandel Q parameter (see Eq.~(\ref{eq:mandel-q})), and the transverse squared-electric field (see Eq.~(\ref{eq:E-field-square})) of their respective gauges for the correlated atom-photon ground-state for different light-matter coupling strengths. That is, we compute $\langle \Psi_{\text{L},0}| \hat{O}_{\text{L}} |\Psi_{\text{L},0} \rangle$ and $\langle \Psi_{\text{V},0}| \hat{O}_{\text{V}}|\Psi_{\text{V},0} \rangle$  where $\hat{O}_{\text{L}}$ and $\hat{O}_{\text{V}}$ are, respectively, observables defined in the length and velocity gauges, and $|\Psi_{\text{L},0} \rangle$ and $|\Psi_{\text{V},0} \rangle$ are the corresponding correlated atom-photon ground-state. To highlight a common misrepresentation of observables often encountered in polaritonic chemistry, we show what happens when observables are incorrectly defined, specifically in the length gauge. That is, the operators are not derived by unitarily transforming their parent velocity-gauge form. We consider the following \textit{incorrect} definitions of the photon number operator, Mandel $Q$ parameter and the squared transverse electric field given respectively as 
\begin{equation}
\begin{aligned}
\hat{n}_{\alpha,\text{L}'} &= \hat{a}_{\alpha,\text{L}'}^{\dagger} \, \hat{a}_{\alpha,\text{L}'} \, , \\
Q_{\alpha,\text{L}'} &=  \frac{\langle \hat{a}_{\alpha,\text{L}'}^{\dagger}\hat{a}_{\alpha,\text{L}'}^{\dagger}\hat{a}_{\alpha,\text{L}'}\hat{a}_{\alpha,\text{L}'}\rangle - \langle \hat{a}_{\alpha,\text{L}'}^{\dagger}\hat{a}_{\alpha,\text{L}'}\rangle^{2} }{\langle \hat{a}_{\alpha,\text{L}'}^{\dagger}\hat{a}_{\alpha,\text{L}'}\rangle} \, , \\
\hat{\textbf{E}}_{\alpha,\text{L}'}^{2} &= \boldsymbol{\lambda}_{\alpha}^{2} \left( \frac{\hbar\omega_{\alpha}}{2} \right) \left[2\hat{a}_{\alpha,\text{L}'}^{\dagger} \, \hat{a}_{\alpha,\text{L}'} + 1 - \hat{a}_{\alpha,\text{L}'}^{\dagger \, 2} - \hat{a}_{\alpha,\text{L}'}^{2} \right] \, .\label{eq:mixed-observables}
\end{aligned}
\end{equation}
Here, the \textit{incorrect} bosonic creation and annihilation operators in Eq.~(\ref{eq:mixed-observables}) are given respectively by
\begin{equation}
\begin{aligned}
\hat{a}_{\alpha,\text{L}'}^{\dagger} &= \frac{1}{\sqrt{2\hbar \omega_{\alpha}}} \left(\omega_{\alpha} \hat{q}_{\alpha,\text{L}'} - i \hat{p}_{\alpha, \text{L}'} \right) , \\
\hat{a}_{\alpha,\text{L}'} &= \frac{1}{\sqrt{2\hbar \omega_{\alpha}}} \left(\omega_{\alpha} \hat{q}_{\alpha, \text{L}'} + i \hat{p}_{\alpha, \text{L}'} \right) . 
\end{aligned} \label{eq:a-adag-LG-incorrect}
\end{equation}
We compute \textit{incorrect} observables as $\langle \Psi_{\text{L},0}| \hat{O}_{\text{L}'}|\Psi_{\text{L},0} \rangle$. For example, the \textit{incorrect} photon mode occupation is computed as $\langle \Psi_{\text{L},0}| \hat{n}_{\alpha,\text{L}'} |\Psi_{\text{L},0} \rangle$. In Fig.~(\ref{fig:gs-observables-LG-VG}) we show a comparison between the results of the velocity gauge (VG) and the length gauge (LG) as well as the incorrect results. The VG and LG agree quantitatively, as expected, for a varying light-matter coupling strength, while the incorrect LG results differ for increasing light-matter coupling strength. This clearly highlights the importance of using the correct observables defined for the Hamiltonian description under consideration. In the results of Fig.~(\ref{fig:gs-observables-LG-VG}), we used an effective cavity mode frequency $\omega_{\alpha}=0.1$~a.u. We note that the results of the VG and LG are always quantitatively the same when we vary the cavity frequency, however, the incorrect LG results always change and differ for increasing light-matter coupling strength.

\section{Model systems of matter}

In this section, we present details of the model systems of the different matter systems considered in this work.

\subsection{One-dimensional model of the hydrogen atom}
\label{sup:1D-atomic-hydrogen}

The model of the hydrogen atom in one dimension is described by the Hamiltonian~\cite{su1991,loudon1959,chen2021}
\begin{align}
\hat{H}_{\text{A}} &= - \frac{\hbar^{2}}{2m_{e}}\frac{\partial^{2} }{\partial x^{2}} -\frac{Ze^{2}}{4\pi\epsilon_{0}}\frac{1}{\sqrt{x^{2} + a_{en}}} \,  , \label{eq:1D-atom}
\end{align}
where we have replaced the electron-nuclear potential by a
soft Coulomb potential, where $a_{en}$ is the softening parameter and $Z$ is the nuclear charge. For our calculations, we represent the bound electron on a uniform real-space grid of $N_{x} = 3000$ grid points with grid spacing $\Delta x = 0.0707$ a.u. while applying an eighth-order finite-difference scheme for the Laplacian. We choose $a_{en}=2$ such that by diagonalizing the static Schr\"{o}dinger for Eq.~(\ref{eq:1D-atom}) gives a ground-state energy $E_{0}=-0.5$~a.u., which is the same as that of the three-dimensional hydrogen atom.

\subsection{One-dimensional model of the hydrogen molecule H$_{2}$}
\label{sup:1D-molecular-hydrogen}

The model of the hydrogen molecule H$_{2}$ in one dimension is given by the Hamiltonian~\cite{lively2021,kreibich2001}
\begin{align}
\hat{H}_{\text{M}} &= - \frac{1}{2\mu_{n}}\frac{\partial^{2} }{\partial R^{2}} + \frac{1}{R} + \frac{1}{\sqrt{(x_{1}-x_{2})^{2} + a_{\text{ee}}}} \nonumber \\
& \quad + \sum_{i=1}^{2} \left(- \frac{1}{2\mu_{\text{e}}}\frac{\partial^{2} }{\partial x_{i}^{2}} -\frac{1}{\sqrt{(x_{i}-R/2)^{2} + a_{\text{en}}}} \right .\\
&\qquad\qquad\quad \left .  - \frac{1}{\sqrt{(x_{i}+R/2)^{2} + a_{\text{en}}}}\right) , \nonumber
\label{eq:1D-H2-hamiltonian}
\end{align}
where $\mu_{\text{e}}=2M_{n}/(2M_{n}+1)$ and $\mu_{n}=M_{n}/2$ are, respectively, the reduced observable electronic and nuclei masses, and the proton mass is $M_{n}= 1836 \, m_{e}$. The electronic coordinates $x_{1}$ and $x_{2}$ describe the two electrons and $R$ is the internuclear separation. The terms of electron-electron and electron-nuclear interaction are represented by soft-Coulomb potentials where the soft-Coulomb parameters take values $a_{\text{ee}}\!=\!2$ and $a_{\text{en}}\!=\!1$. To numerically describe the one-dimensional hydrogen molecule H$_{2}$, we use a grid $(0, 9]$ a.u. for the internuclear separation with a uniform grid spacing $\Delta R = 0.1$ a.u. For the electron degrees,  we represent both electrons on a uniform real-space grid of $N_{x_{1}} = N_{x_{2}}= 200$ grid points with grid spacing $\Delta x_{1} = \Delta x_{2} =  0.35$ a.u. At the nuclear equilibrium position $R_{\text{eq}}=1.9$ a.u., the corresponding ground-state energy is $E_{0}=-1.4843$ a.u.

\subsection{Two-dimensional semiconductor GaAs quantum ring}
\label{sup:quantum-ring}

Our model of a semiconductor quantum ring of GaAs features a single effective electron restricted to two dimensions in real space ($\hat{\textbf{r}} = \hat{x} \textbf{e}_{x} + \hat{y} \textbf{e}_{y}$). The effective electron is confined within a Mexican
hat potential $v_{\text{ext}}(\textbf{r})$ and the Hamiltonian describing the dynamics of the system is given by~\cite{rasanen2007,flick2015,flick2017b,welakuh2021,welakuh2022c}:
\begin{equation}
\hat{H}_{\text{QR}} = -\frac{\hbar^{2}}{2m} \left(\frac{\partial^{2}}{\partial x^{2}} + \frac{\partial^{2}}{\partial y^{2}}\right)  + \underbrace{\frac{1}{2}m\omega_{0}^{2}\hat{\textbf{r}}^{2} + V_{0} e^{-\hat{\textbf{r}}^{2}/d^{2}}}_{v_{\text{ext}}(\textbf{r})} , \label{eq:quantum-ring-hamiltonian}
\end{equation}
The potential parameters are chosen such that they reflect the energy and length scales of a semiconductor quantum ring of GaAs as used in experiments~\cite{fuhrer2001,ihn2005} where $\hbar \omega_{0} = 10$ meV, $d = 10$ nm, $m = 0.067 m_{e}$ and $V_{0} =200$ meV. The Hamiltonian of Eq.~(\ref{eq:quantum-ring-hamiltonian}) is represented on a two-dimensional uniform real-space grid of $N_{x} = N_{y} = 127$ grid points (implying that $127^{2}$ states are considered) with grid spacing $\Delta x = \Delta y = 0.7052$ nm while applying an eighth-order finite-difference scheme for Laplacian operator. 

\begin{figure*}[bth]
\includegraphics[width=6.5in,height=6.0in]{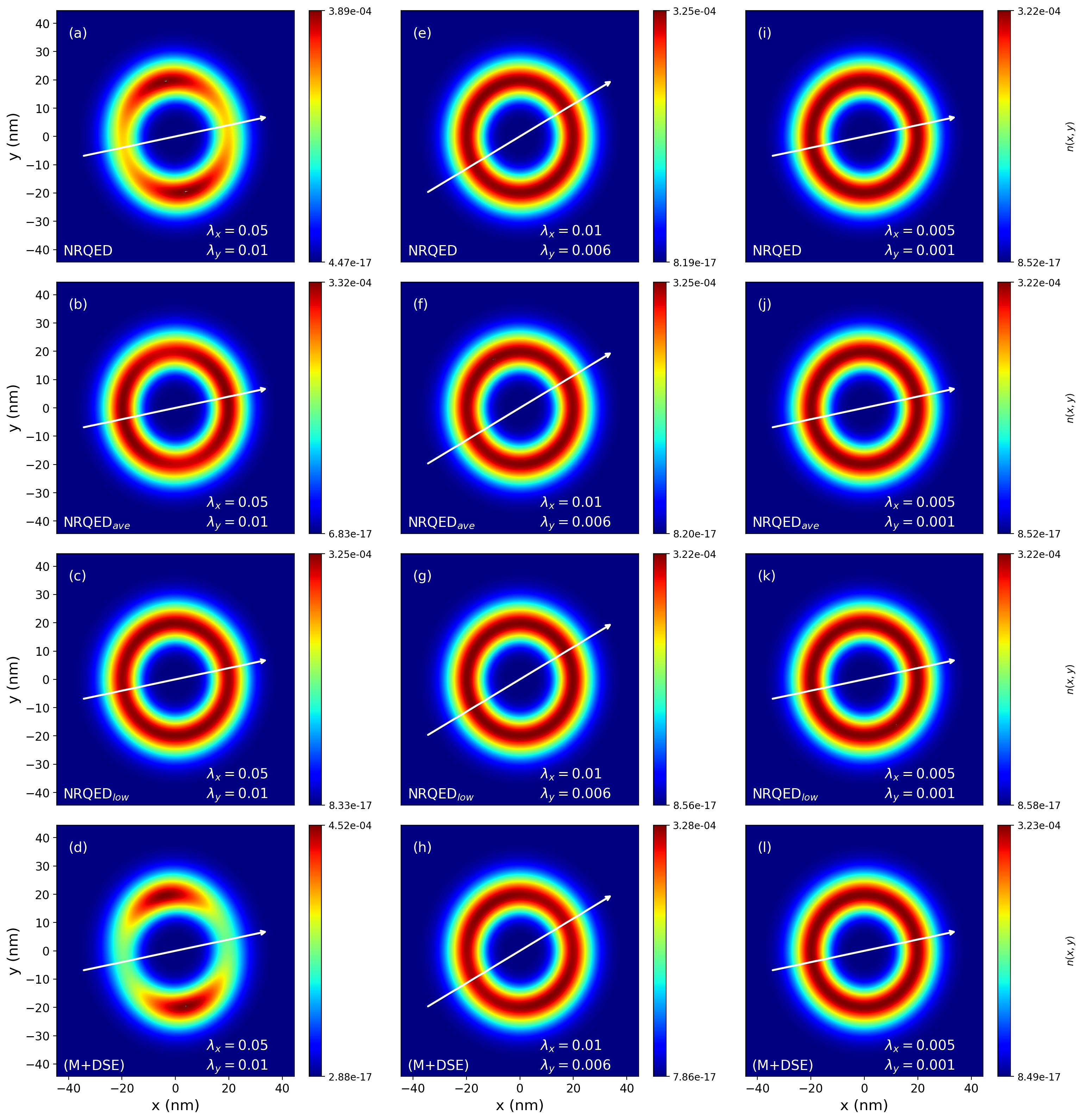}
\caption{The numerical exact ground-state density $n(x,y)$ and the different approximation strategies for different light-matter coupling strengths $\lambda_{x,y}$. For (a)--(d) the couplings are $\lambda_x=0.05$ and $\lambda_y=0.01$. For (e)--(h) the couplings are $\lambda_x=0.01$ and $\lambda_y=0.006$, and for (i)--(l) the couplings are $\lambda_x=0.005$ and $\lambda_y=0.001$. For (a)--(d) we observe a clearly visible breaking of rotational symmetry in the ground-state density.    }
\label{fig:gs-GaAs-exact}
\end{figure*}

In Fig.~(\ref{fig:gs-GaAs-exact}) we show the exact numerical correlated electron-photon ground-state density and the different approximation strategies for different light-matter coupling strengths. In the cases where the coupling $\lambda_y$ is that order $\lambda_y\sim 10^{-3}$ the densities are effectively rotationaly symmetric. In the case however, where $\lambda_y=10^{-2}$ the densities deform and density accumulates perpendicular to the polarization direction. This leads to a breaking of rotational symmetry in the density which is clearly visible in Figs.~\ref{fig:gs-GaAs-exact} (a), (b), and (c).

\begin{figure*}[bth]
\includegraphics[width=6.5in,height=5.0in]{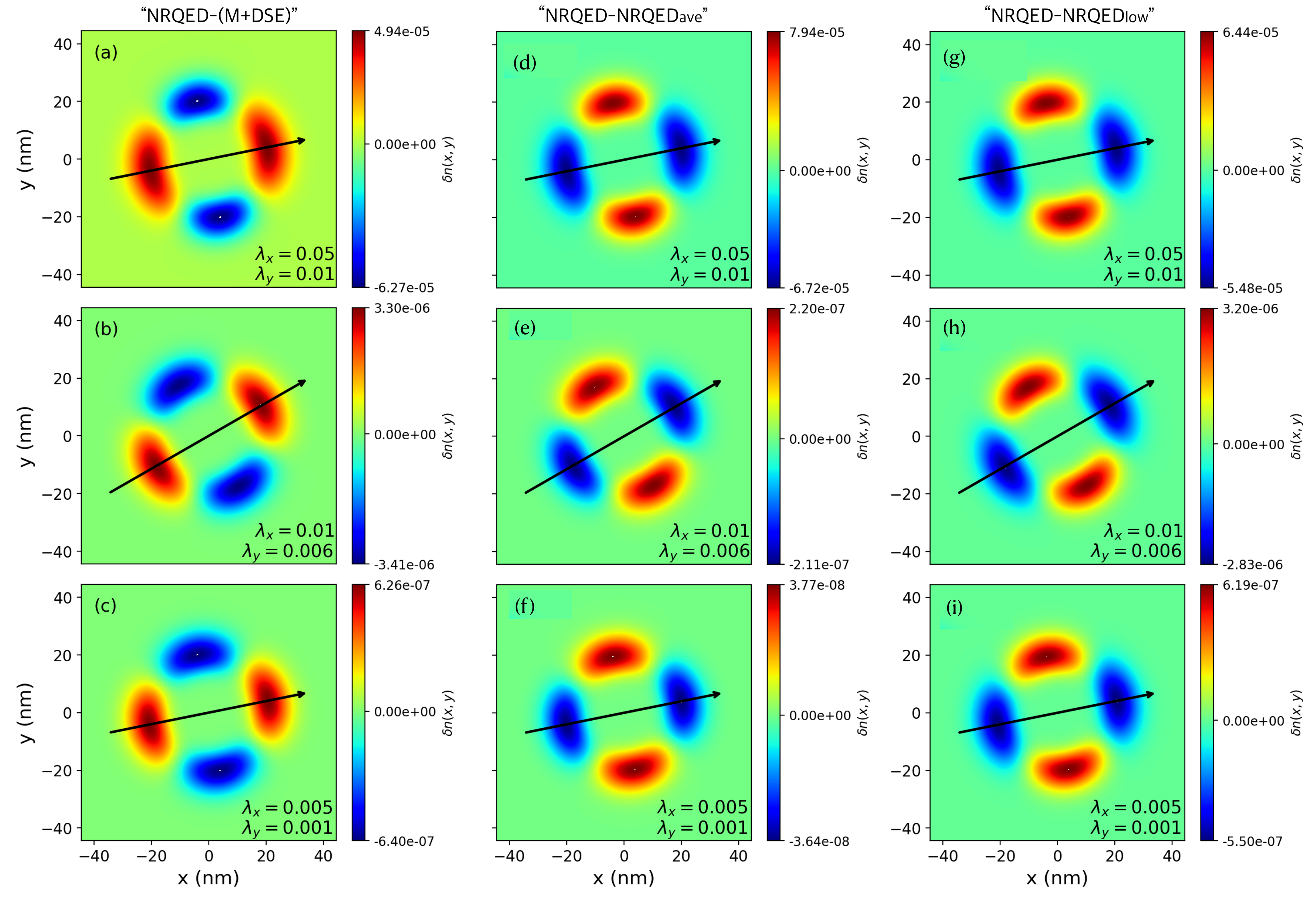}
\caption{The ground-state density difference between the exact NRQED and the different approximation strategies (M+DSE, NRQED$\rm ave$, and NRQED$_{\rm low}$) for the GaAs quantum ring and for different light-matter coupling strengths $\lambda_{x,y}$ along the different columns. For (a)--(c) we show the density difference for M+DSE and varying couplings. For (d)--(f) we show the density difference for NRQED$_{\rm ave}$ and varying couplings. For (g)--(i) we show the density difference for NRQED$_{\rm low}$ and varying couplings. For the couplings $\lambda_x=0.05$ and $\lambda_y=0.01$ all three methods perform effectively the same with deviations at the order $10^{-5}$, with NRQED$_{\rm low}$ only slightly outperforming the other two methods. Overall, we find NRQED$_{\rm ave}$ to be the most accurate approximation for the widest range of light-matter couplings. }
\label{approximation deviations}
\end{figure*}

Furthermore, in Fig.~\ref{approximation deviations} we show the deviations of the each approximation strategy from the exact results for a range of couplings strengths in $x$ and $y$ directions. For most light-matter couplings, $\lambda_x=0.01,0.005$ and $\lambda_y=0.01,0.001$ respectively, the averaging method NRQED$_{\rm ave}$ performs the best as its deviations are one order of magnitude lesser than the other two methods. For the couplings $\lambda_x=0.05$ and $\lambda_y=0.01$ all three methods perform effectively the same with deviations at the order $10^{-5}$, with NRQED$_{\rm low}$ only slightly outperforming the other two methods. Overall, we find NRQED$_{\rm ave}$ to be the most accurate approximation for the widest range of light-matter couplings.

\end{document}